\documentclass[12pt]{iopart}

\usepackage{bm,amstext,subeqn,iopams}  
\usepackage{graphicx}

\newcommand{\eqref}[1]{(\ref{#1})}
\newcommand{\binom}[2]{{#1\choose #2}}
\newcommand{\tfrac}[2]{{\textstyle{\frac{#1}{#2}}}}
\newcommand{\dfrac}[2]{{\displaystyle{\frac{#1}{#2}}}}
\newenvironment{mycases}{\left\{\begin{array}{ll}}{\end{array}\right.}

\newcommand{\RE}{\text{Re}}
\newcommand{\IM}{\text{Im}}
\newcommand{\I}{\text{i}}
\newcommand\floor[1]{\lfloor#1\rfloor}

\newcommand{\ORDER}[1]{\text{O}(#1)}

\newcommand{\ORDERBig}[1]{\text{O}\Big(#1\Big)}

\newcommand{\sx}{\sigma_x}
\newcommand{\sy}{\sigma_y}
\newcommand{\sz}{\sigma_z}
\newcommand{\tx}{\tau_x}
\newcommand{\ty}{\tau_y}
\newcommand{\tz}{\tau_z}
\newcommand{\bra}[1]{\langle{#1}|}
\newcommand{\ket}[1]{|{#1}\rangle}
\newcommand{\braket}[2]{\langle{#1}|{#2}\rangle}
\newcommand{\expval}[1]{\langle{#1}\rangle}
\newcommand{\coh}{{\text{c}}}
\newcommand{\mydag}{\dag}
\newcommand{\phdag}{{\phantom{\mydag}}}
\newcommand{\comm}[2]{\left[{#1},{#2}\right]}

\newcommand{\hypM}{M}
\newcommand{\hypMREG}{\mathbf{M}}
\newcommand{\hyppFq}{{}_p\text{F}_q}
\newcommand{\hypiFi}{{}_1\text{F}_1}

\newcommand{\hypiiFo}{{}_2\text{F}_0}

\newcommand{\hypiiiFii}{{}_3\text{F}_2}
\newcommand{\Beta}{\text{B}}
\newcommand{\NOTILDE}[1]{#1}
\newcommand{\gTILDE}{\tilde{g}}
\newcommand{\dTILDE}{\tilde{\Delta}}
\newcommand{\calC}{{\cal C}}
\newcommand{\calF}{{\cal F}}
\newcommand{\calG}{{\cal G}}

\newcommand{\setZ}{\mathbb{Z}}
\newcommand{\setC}{\mathbb{C}}
\newcommand{\setR}{\mathbb{R}}
\newcommand{\calH}{{\cal H}}
\newcommand{\Hp}{{\cal H}_+}
\newcommand{\Hm}{{\cal H}_-}
\newcommand{\ua}{\uparrow}
\newcommand{\da}{\downarrow}
\newcommand{\mynewpage}{}

\begin{document}

  \title%
  [Asymptotics in the asymmetric quantum Rabi model]%
  {Asymptotic behavior of observables in the asymmetric quantum Rabi model}

  \author{J Semple\footnote{Present address:
      School of Science,
      Applied Science and Management Division,
      Yukon College,
      PO~Box 2799,
      Whitehorse,
      Yukon Y1A~5K4, 
      Canada}
    and
    M Kollar
  }

  \address{Theoretical Physics III, Center for Electronic
    Correlations and Magnetism, Institute of Physics, University of
    Augsburg, 86135~Augsburg, Germany}
  
  \begin{abstract}
    The asymmetric quantum Rabi model with broken parity invariance
    shows spectral degeneracies in the integer case, that is when
    the asymmetry parameter equals an integer multiple of half the
    oscillator frequency, thus hinting at a hidden symmetry and
    accompanying integrability of the model. We study the expectation
    values of spin observables for each eigenstate and observe
    characteristic differences between the integer and noninteger 
    cases for the asymptotics in the deep strong coupling regime,
    which can be understood from a perturbative expansion in the qubit
    splitting. We also construct a parent Hamiltonian whose exact
    eigenstates possess the same symmetries as the perturbative
    eigenstates of the asymmetric quantum Rabi model in the integer case.
  \end{abstract}

  \pacno{42.50.Pq}
  \submitto{\jpa}
  \maketitle  

  \section{\label{sec:introduction}Introduction}

  The importance of the simplest model for light-matter interaction,
  introduced in its semi-classical form by I.~I.~Rabi eighty years ago,
  has been emphasized in the introductory article to this special
  issue~\cite{braak_semi-classical_2016}.
  Its fully quantized version, the quantum Rabi model (RM), given by the
  Hamiltonian
  \begin{eqnarray}
    H_{\text{RM}}
    &=&
    \omega\,a^\mydag a
    +
    g(a^\mydag+a)\,\sx
    +
    \Delta\,\sz
    \,,\label{eq:H_RM}
  \end{eqnarray}
  was studied using the rotating-wave approximation by Jaynes and
  Cummings~\cite{jaynes_comparison_1963}. It involves a spin-$\frac12$
  with standard Pauli matrices $\bm{\sigma}$,
  \begin{eqnarray}
    \sx=\left(\begin{array}{rr} 0& 1\\ 1& 0\end{array}\right),
    ~~~~
    \sy=\left(\begin{array}{rr} 0&-i\\ i& 0\end{array}\right),
    ~~~~
    \sz=\left(\begin{array}{rr} 1& 0\\ 0&-1\end{array}\right),
  \end{eqnarray}
  representing an atomic two-level system (qubit). The qubit splitting
  is given by $\Delta=\Omega/2$, where $\Omega$ represents the qubit
  frequency (we set $\hbar$ $=$ $1$ throughout), or, in solid-state
  realizations such as circuit QED~\cite{niemczyk_circuit_2010} as
  well as in the polaron picture~\cite{ying_ground-state_2015}, the
  energy gap or hybridization. The qubit couples to the radiation
  field (in dipole approximation), which is described by a single
  harmonic oscillator with number eigenstates $\ket{n}$, where
  $a^\mydag a\ket{n}$ $=$ $n\ket{n}$ and $\comm{a}{a^\mydag}$ $=$ $1$.
  Realizations of~\eqref{eq:H_RM} with ultracold atomic gases were
  also proposed recently~\cite{schneeweiss_cold-atom_2017}.

  The Hamiltonian~\eqref{eq:H_RM} has an important discrete symmetry,
  because it commutes with the operator $P$ $=$
  $\sz(-1)^{a^\mydag a}$. Since $P^2$ $=$ $1$, the symmetry group
  is $\setZ_2$, i.e.,  the Hilbert space 
  $\calH$ $=$ $L^2(\setR)\otimes\setC^2$ is the direct sum of
  two invariant spaces with fixed parity, $\calH$ $=$ $\Hp\oplus\Hm$,
  \begin{subequations}
    \begin{eqnarray}
      \calH_p
      &=&
      \left\{
        \ket{\psi_p}
        \,\big|\,
        P\ket{\psi_P}
        =
        p\ket{\psi_p}
      \right\}
      \,,~~~~p=\pm1
      \,.
    \end{eqnarray}
  \end{subequations}
  This discrete symmetry renders $H_{\text{RM}}$
  integrable~\cite{braak_integrability_2011}.  The rotating-wave
  approximation for $H_{\text{RM}}$ keeps only the coupling of
  $\sigma_\pm$ $=$ $\sx$ $\pm$ $\I\sy$ to $a$ and $a^\mydag$,
  respectively. This extends the symmetry to a continuous
  $U(1)$~symmetry, leading to the superintegrability of the
  Jaynes-Cummings model,
  \begin{eqnarray}
    H_{\text{JCM}}
    &=&
    \omega\,a^\mydag a
    +
    g\,(a^\mydag\,\sigma_-+a\,\sigma_+)
    +
    \Delta\,\sz
    \,,\label{eq:H_JCM}
  \end{eqnarray}
  for which the Hilbert space separates into an infinite number of
  two-dimensional invariant subspaces. In the spectral graph, i.e., a
  plot of eigenenergies as a function of a parameter such as $g$, the
  larger symmetry group of $H_{\text{JCM}}$ creates additional level
  crossings, producing infinitely many two-legged `ladders'; for
  $H_{\text{RM}}$, on the other hand, the spectral graph consists of
  two ladders with infinitely many legs, since there are only two
  invariant subspaces (each infinite dimensional) related to the two
  eigenvalues $p$ $=$ $\pm1$ of
  $P$~\cite{braak_integrability_2011,chen_exact_2012}. These two
  ladders intersect in the spectral graph at the so-called Juddian
  points~\cite{judd_jahnteller_1977}, corresponding to quasi-exact,
  doubly degenerate eigenvalues of
  $H_{\text{RM}}$~\cite{kus_exact_1986,wakayama_quantum_2014}.

  A generalization of $H_{\text{RM}}$ is the {\it asymmetric} quantum
  Rabi model (ARM)~\cite{braak_integrability_2011,xie_anisotropic_2014}
  with Hamiltonian
  \begin{eqnarray}
    H_{\text{ARM}}
    &=&
    H_{\text{RM}}
    +
    \epsilon\,\sx
    =
    \omega\,a^\mydag a
    +
    g(a^\mydag+a)\sx
    +
    \epsilon\,\sx
    +
    \Delta\,\sz
    \,.\label{eq:H_ARM}
  \end{eqnarray}
  The term~$\epsilon\,\sx$ corresponds to spontaneous flips of the
  two-level system and appears naturally in implementations with flux
  qubits~\cite{niemczyk_circuit_2010,manucharyan_resilience_2017}. This term breaks the
  $\setZ_2$~invariance of $H_{\text{RM}}$, so that $H_{\text{ARM}}$
  possesses no obvious symmetry and no level crossings should be
  expected in the spectral graph for any~$\epsilon\neq0$. Indeed, no
  level crossings are observed if~$\epsilon$ is not an integer
  multiple of $\omega/2$, which we call the noninteger case
  henceforth. Furthermore the ARM has a quasi-exact exceptional
  spectrum just like the RM~\cite{li_algebraic_2015}, but the
  quasi-exact eigenstates are no longer doubly degenerate in the 
  noninteger case.

  Surprisingly, degeneracies are observed again if~$\epsilon$ is an
  integer multiple of $\omega/2$, to which we refer as integer values or the integer case
  from now on. These degeneracies correspond to two intersecting ladders
  with infinitely many legs in the spectral graph, just as in the RM,
  thus suggesting the presence of a hidden $\setZ_2$~symmetry of the
  ARM in the integer case~\cite{braak_integrability_2011,batchelor_integrability_2015}.
  This symmetry would make the integer case of the ARM (iARM) integrable
  again according to the level labeling criterion proposed in
  \cite{braak_integrability_2011}. For this reason, the ARM has been
  under intense
  study~\cite{li_algebraic_2015,li_addendum_2016,batchelor_energy_2016,xie_quantum_2017}. In
  fact, a proof for these degeneracies was developed for the case
  $\epsilon$ $=$
  $\omega/2$~\cite{wakayama_quantum_2014,wakayama_symmetry_2017,reyesbustos_2018}
  and can be generalized to higher half-integer multiples of $\omega$.

  In this paper, we investigate another aspect of the ARM, namely the
  effect of integer values of the asymmetry parameter on the asymptotics of spin
  expectation values in the energy eigenstates, which were also
  recently studied in~\cite{meher_academia_2015}.  We begin with
  numerical results for these quantities in
  Sec.~\ref{sec:numerical}. These can be understood perturbatively for
  small hybridization $\Delta$ (Secs.~\ref{sec:perturbative},
  \ref{sec:inco-pt}, \ref{sec:comm-pt}).  Furthermore, the
  lowest-order perturbative eigenstates serve as eigenstates of a
  parent Hamiltonian which we construct in
  Sec.~\ref{sec:parenthamiltonian}, and which might be useful in
  understanding the symmetries of the iARM. We close with a summary
  and outlook in Sec.~\ref{sec:conclusion}.

  \section{\label{sec:numerical}Spin expectation values}

  In this section, we discuss numerical results for spin expectation
  values in the energy eigenstates of the ARM. For a general state,
  \begin{eqnarray}
    \ket{\psi}
    &=&
    \left(
      \begin{array}{c}
        \ket{\phi_{\ua}}\\
        \ket{\phi_{\da}}
      \end{array}
    \right)\,,
  \end{eqnarray}
  where $\ket{\phi_\sigma}$ are states of the oscillator, expectation
  values of the spin operator are given by
  \begin{subequations}
    \begin{eqnarray}
      \expval{\sx} &=& 2\,\RE\,\braket{\phi_\ua}{\phi_\da}
      \,,\\
      \expval{\sy} &=& 2\,\IM\,\braket{\phi_\ua}{\phi_\da}
      \,,\\
      \expval{\sz} &=& \braket{\phi_\ua}{\phi_\ua} - \braket{\phi_\da}{\phi_\da}
      \,.
    \end{eqnarray}
  \end{subequations}

  \subsection{Symmetric Rabi model}

  Consider first the symmetric RM, i.e., the case $\epsilon$ $=$ $0$, 
  All eigenstates $\ket{\psi_p}$ with fixed parity $p=\pm1$ have the form
  \begin{equation}
    \ket{\psi_+}=\left(
      \begin{array}{c}
        \ket{\phi_{\ua-}}^{\text{e}}\\
        \ket{\phi_{\da-}}^{\text{o}}
      \end{array}
    \right)\,,
    \qquad
    \ket{\psi_-}=\left(
      \begin{array}{c}
        \ket{\phi_{\ua-}}^{\text{o}}\\
        \ket{\phi_{\da-}}^{\text{e}}
      \end{array}
    \right)\,,
    \label{eq:even-odd}
  \end{equation}
  where 
  \begin{equation}
    \ket{\phi^{\text{e}}}=\sum_{n=0,2,4,\ldots}\alpha^{\text{e}}_n\ket{n}
    \,,\qquad
    \ket{\phi^{\text{o}}}=\sum_{n=1,3,5,\ldots}\alpha^{\text{o}}_n\ket{n}
    \,.
    \label{eq:parity}
  \end{equation}
  Therefore the overlap $\braket{\phi_\ua}{\phi_\da}$ which appears in
  $\expval{\sigma_{x}}$, $\expval{\sigma_{y}}$ vanishes for all
  eigenstates of $H_{\text{RM}}$, because Fock states with an even
  number of photons are orthogonal to all states with an odd
  number. Hence $\expval{\sx}$ and $\expval{\sy}$ are zero for the RM,
  and in particular independent of $g$ and $\Delta$.

  \subsection{Asymmetric Rabi model}

  Consider next the ARM with $\epsilon$ $\neq$ $0$. In this case the
  parity invariance is broken and the value of $\expval{\sx}$ and
  $\expval{\sz}$ depend on the state, or, for the energy eigenstates,
  on $g$ and $\Delta$. ($\expval{\sy}$ remains identically zero,
  because $H$ can be represented as an orthogonal matrix with real
  eigenvectors.) From now on we set
  \begin{eqnarray}
    \label{eq:Mdef}
    \epsilon&=&\frac12M\omega\,,
  \end{eqnarray}
  so that the integer (noninteger) case corresponds to  integer (noninteger) $M$, respectively.

\begin{figure}[tb]
    \centering
    \includegraphics[width=0.33\textwidth]{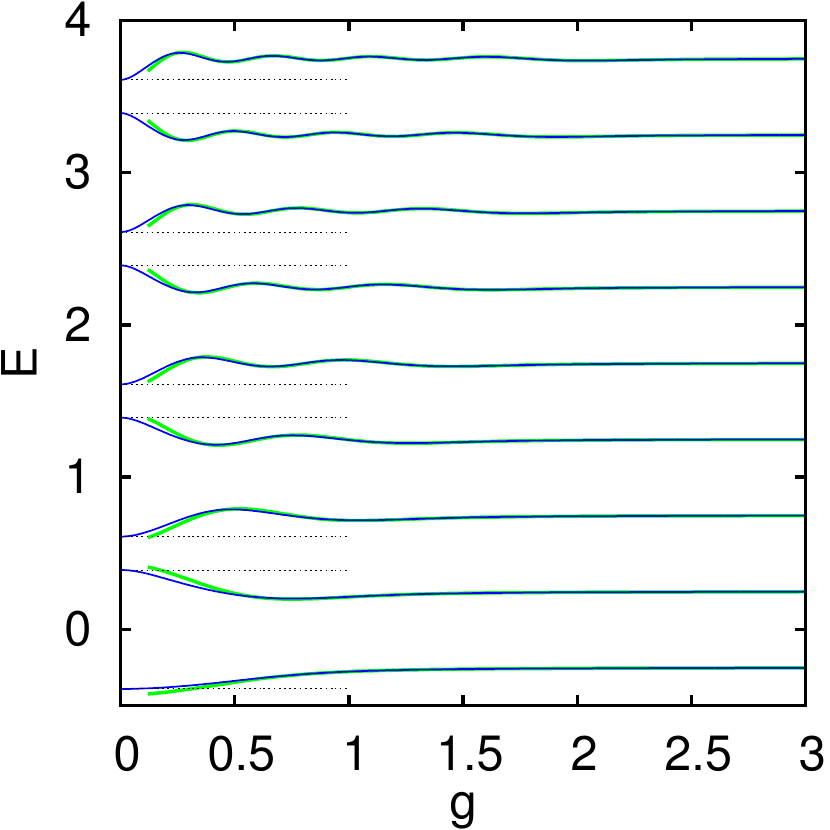}\hfill%
    \includegraphics[width=0.33\textwidth]{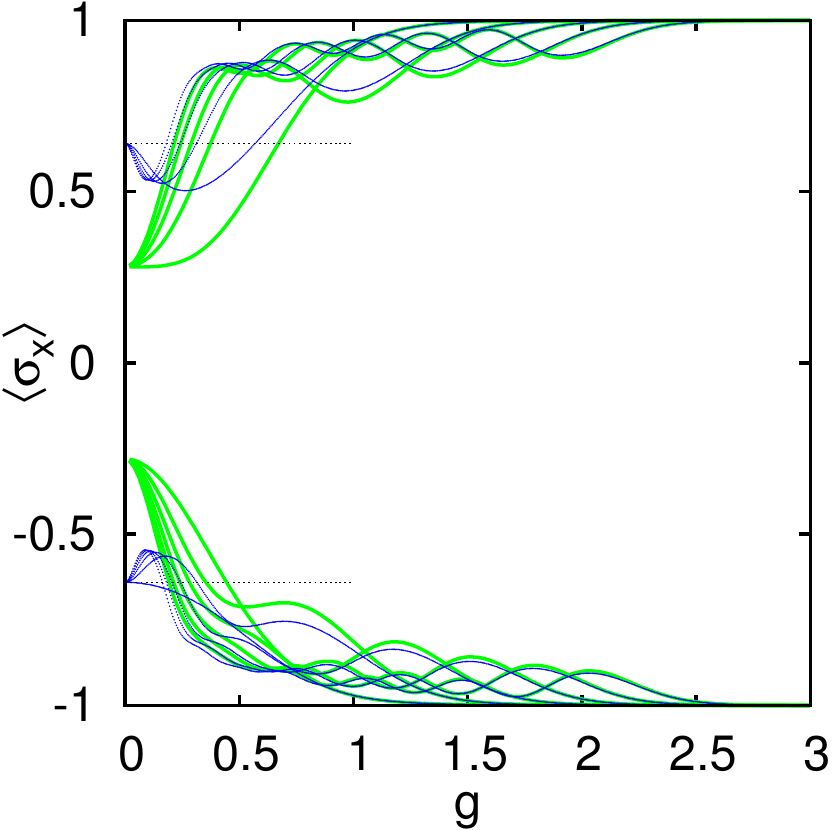}\hfill%
    \includegraphics[width=0.33\textwidth]{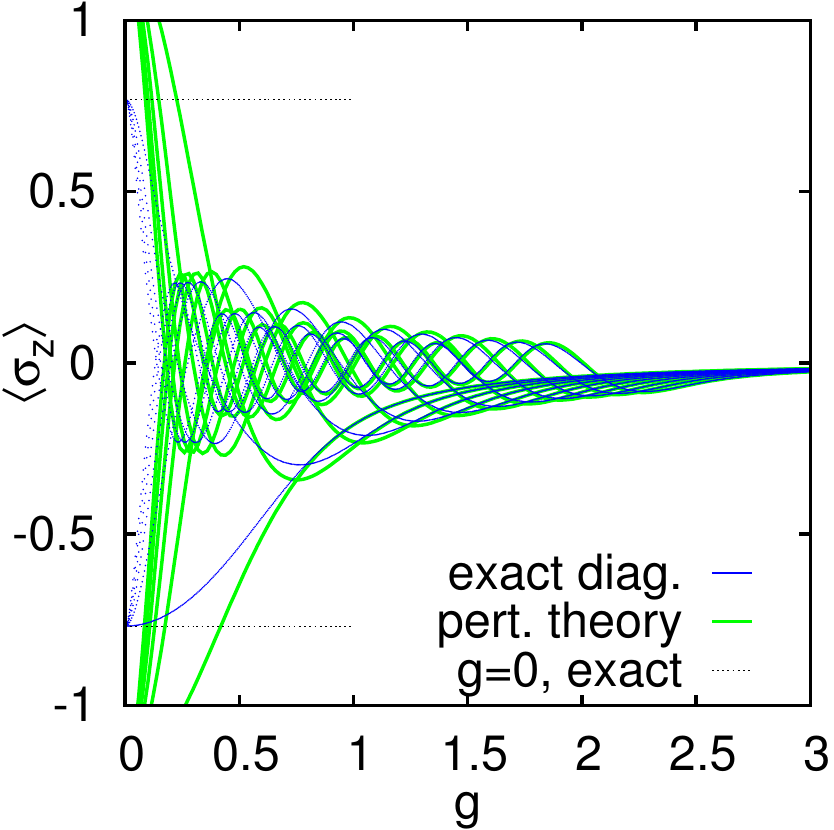}
    \caption{\label{fig1}Eigenenergies and (left) and spin expectation values
      (middle and right) for the noninteger case of the ARM ($M=0.5$, $\epsilon$ $=$
      $0.25$) with small hybridization parameter ($\Delta$ $=$ $0.3$)
      in terms of $\omega$ $=$ $1$ as energy unit. As discussed in the
      text, low-order perturbation theory in $\Delta$ (light/green)
      describes the exact diagonalization results (dark/blue) well
      provided $g$ is not too small. For large $g$, $|\expval{\sx}|$
      tends to 1 and $\expval{\sz}$ tends to 0 for all eigenstates.
      The exact values for $g$ $=$ $0$, given in~\eqref{eq:ARMzerog},
      are marked by dotted lines.}
  \end{figure}

\begin{figure}[p]
    \centering
    \includegraphics[width=0.33\textwidth]{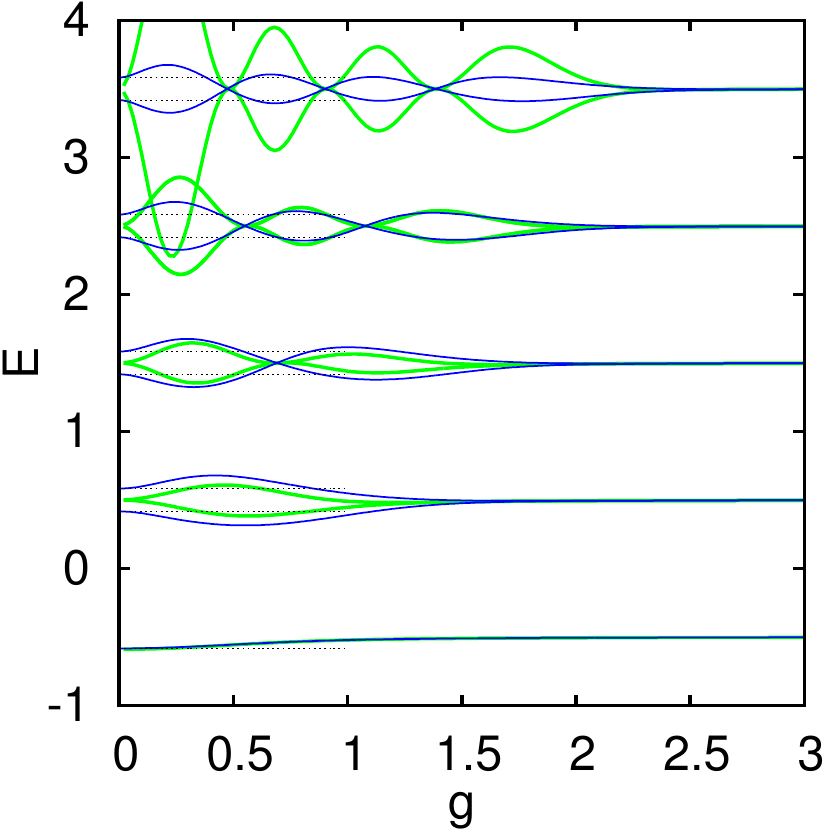}\hfill%
    \includegraphics[width=0.33\textwidth]{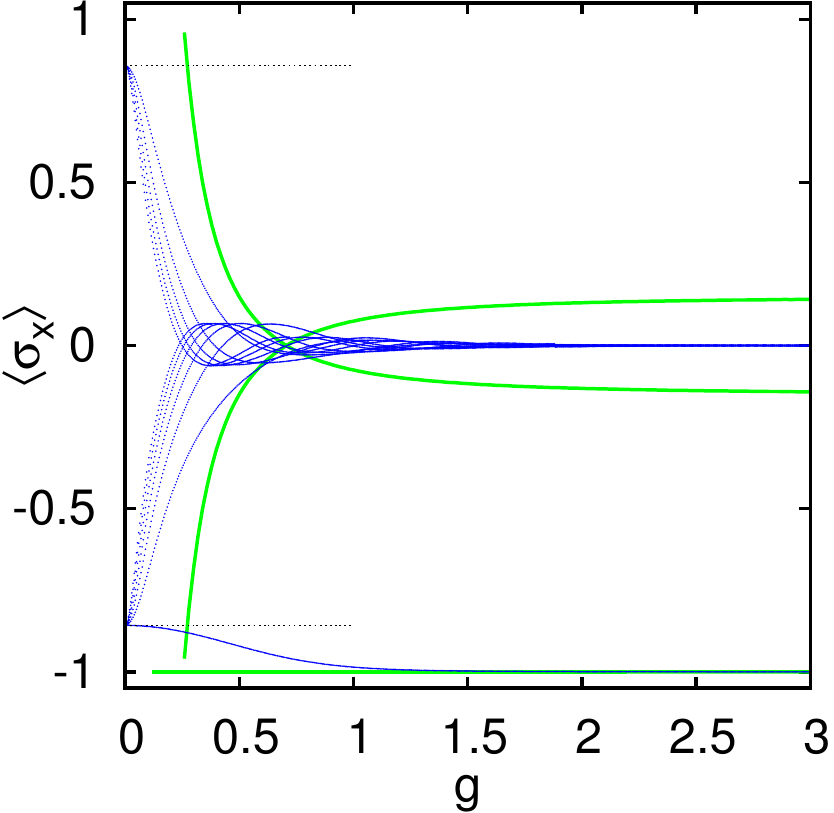}\hfill%
    \includegraphics[width=0.33\textwidth]{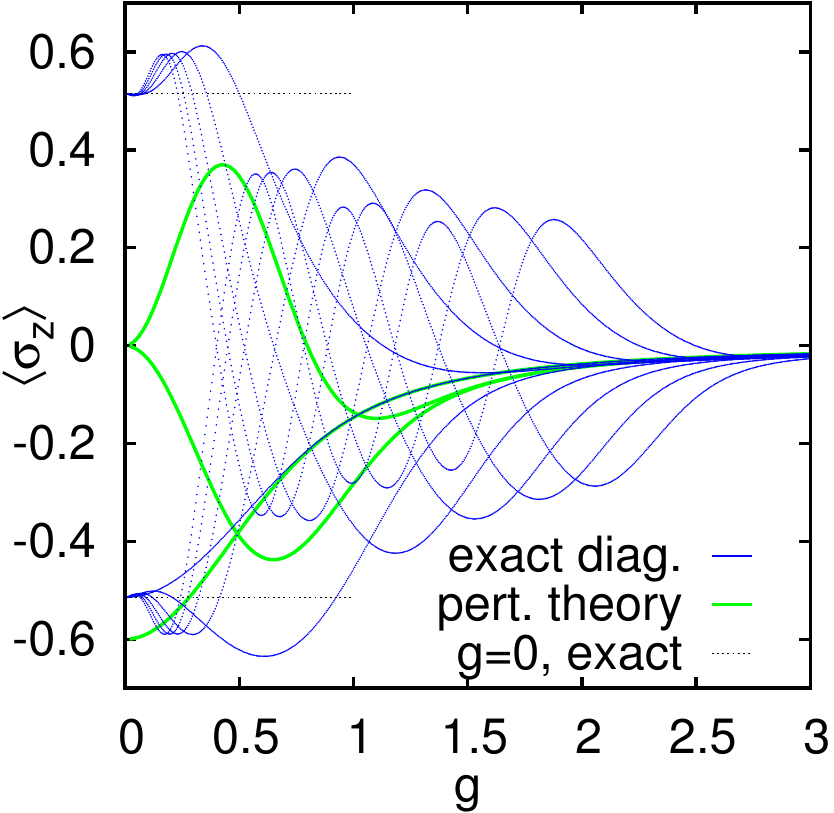}
    \caption{\label{fig2}As Fig.~\ref{fig1}, but for the integer case of the
      ARM (iARM) with $M$ $=$ 1, $\epsilon$ $=$ $0.5$, $\Delta$ $=$ $0.3$,
      $\omega$ $=$ $1$. In the iARM both $|\expval{\sx}|$ and
      $|\expval{\sz}|$ tend to 0 for large $g$, except for $M$ states
      (i.e., one state in the present case) for which $\expval{\sx}$
      tends to $-1$. The perturbative results for the spin expectation
      values are shown only for the $M+1$ lowest energy eigenstates.}
  \end{figure}

\begin{figure}[p]
    \centering
    \includegraphics[width=0.33\textwidth]{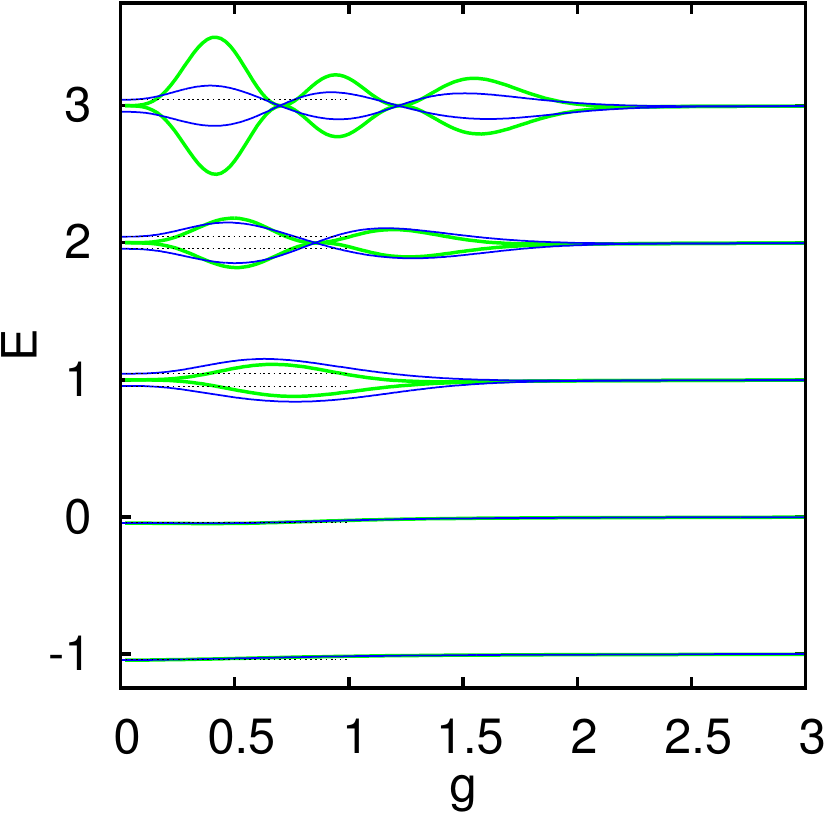}\hfill%
    \includegraphics[width=0.33\textwidth]{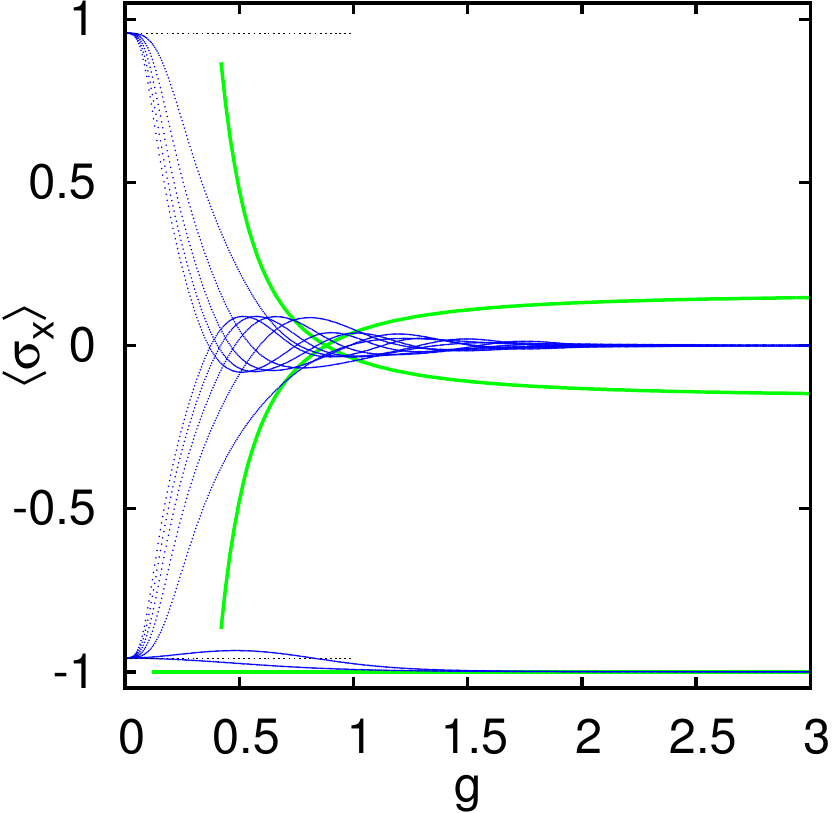}\hfill%
    \includegraphics[width=0.33\textwidth]{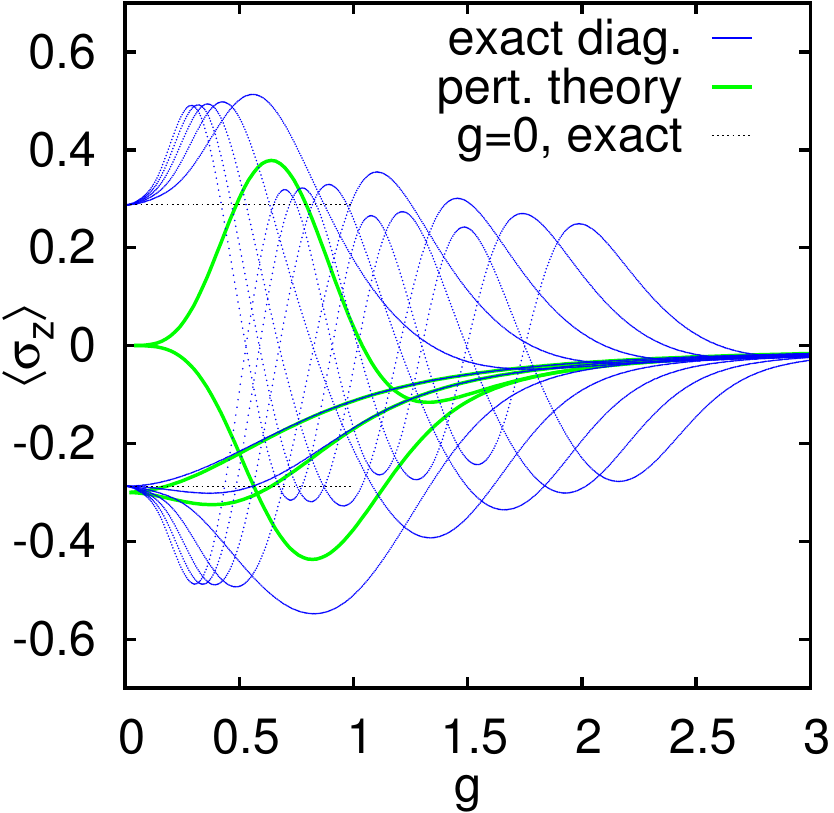}
    \caption{\label{fig3}As Fig.~\ref{fig2} for the iARM, but with $M$ $=$
      2, $\epsilon$ $=$ $1$, $\Delta$ $=$ $0.3$, $\omega$ $=$
      $1$, with two states for which $\expval{\sx}$ tends to $-1$.
    }
  \end{figure}

\begin{figure}[p]
    \centering
    \includegraphics[width=0.33\textwidth]{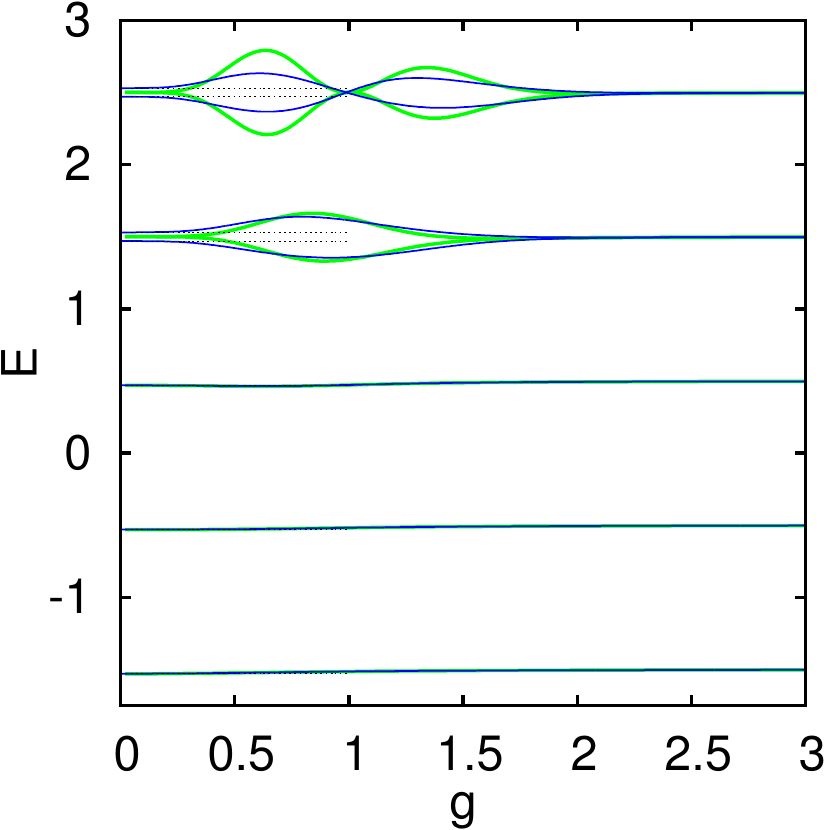}\hfill%
    \includegraphics[width=0.33\textwidth]{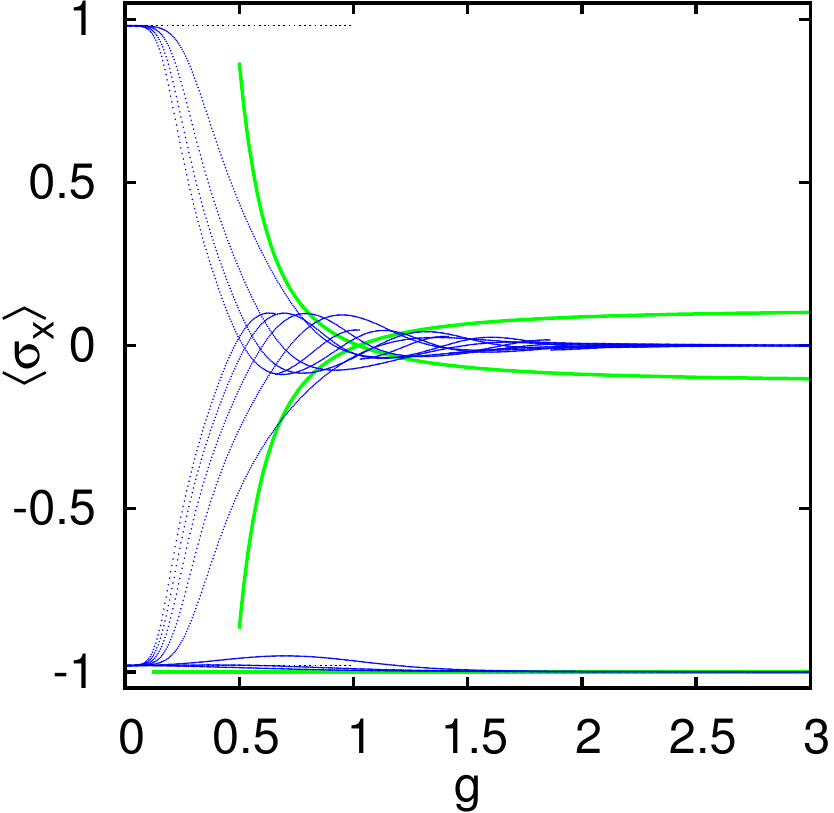}\hfill%
    \includegraphics[width=0.33\textwidth]{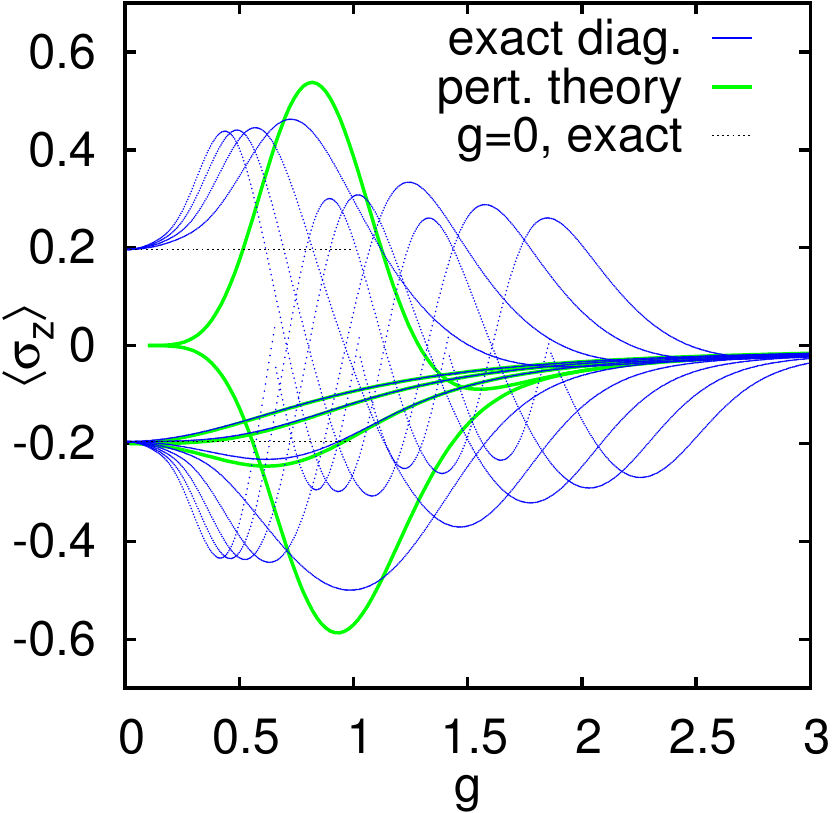}
    \caption{\label{fig4}As Fig.~\ref{fig2}-\ref{fig3} for the iARM, but with $M$ $=$
      3, $\epsilon$ $=$ $1.5$, $\Delta$ $=$ $0.3$, $\omega$ $=$
      $1$, with three states for which $\expval{\sx}$ tends to $-1$.
    }
  \end{figure}

  Figs.~\ref{fig1}-\ref{fig4} show numerical exact diagonalization
  data for $\expval{\sx}$ and $\expval{\sz}$ as a function of $g$ for
  the lowest few eigenstates.  For large $g$, we observe that 
  for integer $M$ the expectation value $\expval{\sx}$ tends to $-1$
  only for the lowest $M$ energy eigenstates (Figs.~\ref{fig2}
  and~\ref{fig3}) and to zero otherwise, while in the noninteger
  case $\expval{\sx}$ tends to $\pm1$. On the other hand,
  $\expval{\sz}$ tends to zero always, for integer and noninteger $M$.

  In the following sections, we offer two `physical' explanations for
  this behavior. On the one hand, we will show that these asymptotics
  are characteristic for small hybridization in the deep strong
  coupling limit (see~\eqref{eq:validity_comm} below), using
  perturbation theory for small hybridization
  (Secs.~\ref{sec:perturbative}-\ref{sec:comm-pt}).
  On the other hand we construct a a related
  Hamiltonian $H'$ with similar properties
  (Sec.~\ref{sec:parenthamiltonian}).

  \mynewpage

  \section{Perturbation theory for weak hybridization $\Delta$: Preliminaries}\label{sec:perturbative}

  In this section we deal with some preparations regarding the
  perturbation theory for small values of $\Delta$.  From now on the
  real parameters~$g$, $\epsilon$, $\Delta$ are assumed to be
  nonnegative without loss of generality. Moreover, we assume $M>0$
  (see~\eqref{eq:Mdef}) from now on, omitting the case of the
  symmetric RM.  The positive energy scale~$\omega$ is often set to
  unity, but we retain it for later convenience. Below we will also
  use the notation $\gTILDE$ $=$ $g/\omega$ and $\dTILDE$ $=$
  $\Delta/\omega$.  

  \subsection{Rotation of spin quantization axis}
  For small $\Delta$ it is preferable to work in the familiar spin-boson picture, i.e., in the eigenbasis
  of~$\sx$. As usual we perform a rotation of the spin quantization axis
  by means of a unitary transformation~$\widetilde{U}$,
  \begin{subequations}
    \begin{eqnarray}
      \widetilde{U}&=&\frac{1}{\sqrt{2}}\left(\begin{array}{rr} 1& 1\\ 1&-1\end{array}\right),
      \\
      \widetilde{\bm{\sigma}}&=&\widetilde{U}^\mydag\bm{\sigma}\widetilde{U}\,,
      ~~~~
      \widetilde{\sigma}_x=\left(\begin{array}{rr} 1&\phantom{-}0\\ 0&-1\end{array}\right)=\tz\,,
      \\
      \widetilde{\sigma}_y&=&\left(\begin{array}{rr} 0&\phantom{-}i\\-i&0\end{array}\right)=-\ty\,,
      ~~~~
      \widetilde{\sigma}_z=\left(\begin{array}{rr} 0&\phantom{-}1\\ 1& 0\end{array}\right)=\tx\,.    
    \end{eqnarray}    
  \end{subequations}
  Here $\bm{\tau}$ are again the standard Pauli matrices, relabelled
  in order to remind us of the transformed basis. In this basis the
  Hamiltonian reads (removing a constant energy term $-g^2/\omega$ and
  omitting the tildes on the transformed Hamiltonian $\widetilde{H}$ and its
  eigenstates, as well as on the the observables $\widetilde{\bm{\sigma}}$),
  \begin{subequations}%
    \label{eq:H}%
    \begin{eqnarray}%
      H&=&\widetilde{U}^\mydag\,H_{\text{ARM}}\,\widetilde{U}+\frac{g^2}{\omega}=H_0+V
      \,,\\[1ex]
      H_0&=&\left(\begin{array}{cc} h_+&      0\\       0& h_-\end{array}\right),
      ~~~~
      V   =\left(\begin{array}{cc}   0& \Delta\\ \Delta&
          0\end{array}\right)=\Delta\,\tau_x\,,
      \\[1ex]
      h_\sigma
      &=&
      \omega\,
      a_\sigma^\mydag
      a_\sigma^\phdag
      +\frac{\sigma M}{2}\,,
      ~~
      a_\sigma^\phdag
      =
      a
      +
      \sigma\gTILDE\,,
      ~~
      \sigma=\pm1
      \,,
    \end{eqnarray}%
  \end{subequations}%
  i.e., $H_0$ can be written in terms of oscillators that are shifted
  in opposite directions.

  \subsection{Unperturbed eigenstates and their overlaps}

  The eigenstates of $H_0$ are
  \begin{eqnarray}
    \ket{n\sigma^{(0)}}
    &=&
    \left(\begin{array}{cc}
        \delta_{\sigma+}\\[1ex]
        \delta_{\sigma-}
      \end{array}\right)
    \ket{n}_\sigma
    \,,
    ~~    
    n=0,1,\ldots,
    ~~
    \sigma=\pm1
    \,,
    \label{eq:unperturbed_eigenstates}
  \end{eqnarray}
  with eigenvalues
  \begin{eqnarray}
    E_{n\sigma}^{(0)}
    &=&
    \omega
    \left(n+\frac{\sigma M}{2}\right)
    \,,
  \end{eqnarray}
  which are nondegenerate for noninteger $M$, whereas for integer $M$
  only the first $M$ energy levels $E_{n-}^{(0)}$ ($n$ $=$
  $0,\ldots M-1$) are nondegenerate, followed by the doubly degenerate
  energy levels $E_{n}^{(0)}$ $=$ $E_{n-M+}^{(0)}$ $=$ $E_{n-}^{(0)}$ ($n$ $=$ $M,M+1,\ldots$).

  In~\eqref{eq:unperturbed_eigenstates} we denoted the eigenstates of
  the shifted number operators $a_\sigma^\mydag a_\sigma^\phdag$ by
  $\ket{n}_{\sigma}$. These can be written in terms of the operators and
  coherent states $\ket{z}_\coh$ of the unshifted oscillator
  (denoting $\bar{\sigma}$ $=$ $-\sigma$) as
  \begin{subequations}
    \begin{eqnarray}
      \ket{n}_\sigma
      &=&
      \frac{
        (a^++\sigma\gTILDE)^n
      }{
        \sqrt{n!}
      }
      \,
      \ket{\bar{\sigma}\gTILDE}_\coh
      \,,
      \\
      \ket{z}_\coh
      &=&e^{za^\mydag}\ket{0}\,e^{-\frac12|z|^2}\,,
      ~~~~
      z\in\mathbb{C}
      \,,
    \end{eqnarray}
  \end{subequations}
  due to the property $a\ket{z}_\coh$ $=$ $z\ket{z}_\coh$ of the
  latter. Below we will encounter the overlap
  \begin{eqnarray}
    \label{eq:def_overlap_F}
    {}_-\braket{n'}{n}_+
    &=&
    F_{n'n}(2\gTILDE)
    \,,
  \end{eqnarray}
  which defines a function $F_{n'n}(x)$ which we now
  calculate. We first introduce shifted coherent states, which are
  related to unshifted coherent states by a translation in $z$,
   \begin{eqnarray}
    \ket{z}_{\coh,\sigma}
    &=&
    e^{-|z|^2/2}
    e^{za_\sigma^\mydag}
    \ket{0}_{\sigma}
    \nonumber\\&=&%=
    \ket{z-\sigma\gTILDE}_{\coh}
    \,
    e^{i\sigma\gTILDE\,\IM z}
    \,.
  \end{eqnarray}
  For shifts in the same direction $\sigma$ they have the usual
  nonzero overlap of coherent states,
  \begin{eqnarray}
    \label{eq:overlaps_shifted_coh_same}
    {}_{\coh,\sigma}\braket{z'}{z}_{\coh,\sigma}
    &=&
    e^{
      -\frac12|z'|^2
      -\frac12|z|^2
      +z'{}^*z
    }
    \,,
  \end{eqnarray}
  while for opposite shifts we obtain
  \begin{eqnarray}
    \label{eq:overlaps_shifted_coh_opposite}
    {}_{\coh,-}\braket{z'}{z}_{\coh,+}
    &=&
    e^{
      -\frac12|z'|^2
      -\frac12|z|^2
      +z'{}^*z
      +2\gTILDE(z-z'{}^*)-2\gTILDE^2
    }
    \\
    &=&
    \sum_{n,n'=0}^{\infty}
    \frac{{}_{-}\braket{n'}{n}_{+}}{\sqrt{n'!n!}}
    (-z'{}^*)^{n'}z^n
    e^{
      -\frac12|z'|^2
      -\frac12|z|^2
      }
    \,.\nonumber
  \end{eqnarray}
  It is useful to work with the following two-variable Hermite
  polynomials $H_{nm}(x,y)$ and their generating
  function~\cite{fan_generating_2015},
  \begin{eqnarray}
    H_{nm}(x,y)
    &=&
    \sum_{k\geq0}
    \binom{n}{k}
    \binom{m}{k}
    k!(-1)^kx^{n-k}y^{m-k}
    \,,
    \\
    e^{-uv+ux+vy}
    &=&
    \sum_{n,m=0}^{\infty}
    \frac{u^nv^m}{n!m!}H_{nm}(x,y)
    \,.\label{eq:Hnm-genfun}
  \end{eqnarray}
  Taking coefficients in~\eqref{eq:overlaps_shifted_coh_opposite}, we
  thus obtain
  \begin{subequations}%
    \label{eq:result_overlap_F}%
    \begin{eqnarray}%
      F_{n'n}(x)
      &=&
      \frac{(-1)^{n'}}{\sqrt{n'!n!}}\,H_{nn'}(x,x)\,e^{-x^2/2}
      \,,
      \\
      {}_{\bar{\sigma}}\braket{n'}{n}_{\sigma}
      &=&
      F_{n'n}(2\gTILDE)\,\sigma^{n+n'}
      \,,~~~~~
      \sigma=\pm1
      \,.
    \end{eqnarray}
  \end{subequations}
  We note that for equal arguments the two-variable
  Hermite polynomials can be written as 
  \begin{subequations}%
    \begin{eqnarray}%
      H_{nm}(x,x)
      &=&
      x^{n+m}\hypiiFo(-n,-m;-\tfrac{1}{x^2})
      \\
      &=&
      (-1)^nx^{m-n}L_n^{(m-n)}(x^2)
      \,,
      ~~m\geq n
      \,,
    \end{eqnarray}
  \end{subequations}
  where $L_n^{(\alpha)}(x)$ are associated Laguerre polynomials, and
  hypergeometric functions are defined as
  \begin{eqnarray}
    \hyppFq(a_1,\ldots,a_p;b_1,\ldots,b_q;z) 
    =
    \sum_{n=0}^{\infty}
    \frac{\prod_{j=1}^p(a_j)_n}{\prod_{j=1}^q(b_j)_n}\frac{z^n}{n!}
    \,,
  \end{eqnarray}
  with Pochhammer symbols expressing rising factorials, i.e., $(a)_n$
  $=$ $\Gamma(a+n)/\Gamma(a)$ in terms of the Euler Gamma function.

  \subsection{A priori conditions for the validity of perturbation theory}
  
  We expect that $V$ may be treated perturbatively if $\Delta$ is
  sufficiently small, i.e. small compared to energy differences of the
  unperturbed energy eigenstates, which involve $\omega$ and possibly
  $\epsilon$.  Another energy scale in the problem is $g$, so that we
  must necessarily require $g$ $\gtrsim$ $\omega$ in order to be able
  to expand in $\Delta/\omega$, and not having to expand in $g/\Delta$
  first. This puts us in the deep strong coupling regime~\cite{rossatto_spectral_2017},
  \begin{eqnarray}
    \Delta \ll \omega\lesssim g
    \,.\label{eq:validity_comm}
  \end{eqnarray}
  As an opposite point of reference, we note the spectrum and
  expectation values when harmonic oscillator and qubit decouple for $g$ $=$ $0$,
  \begin{subequations}%
    \label{eq:ARMzerog}%
    \begin{eqnarray}
      E_{n\sigma}^{(g=0)}
      &=&
      \omega n
      +
      \sigma\,
      \sqrt{\epsilon^2+\Delta^2}
      \,,~~~~
      \sigma=\pm1
      \,,
      \\
      \ket{n\sigma}^{(g=0)}
      &=&
      \frac{1}{\sqrt{2}\,\big(\epsilon^2+\Delta^2+\sigma\Delta\sqrt{\epsilon^2+\Delta^2}\big)^{\frac12}}
      \left(
        \begin{array}{c}
          \Delta+\sigma\sqrt{\epsilon^2+\Delta^2}\\
          \epsilon
        \end{array}
      \right),
      \\[1ex]
      \expval{\sx}_{n\sigma}^{(g=0)}
      &=&
      \frac{\sigma\,\epsilon}{\sqrt{\epsilon^2+\Delta^2}}
      \,,
      ~~~~%\\
      \expval{\sz}_{n\sigma}^{(g=0)}
      =
      \frac{\sigma\,\Delta}{\sqrt{\epsilon^2+\Delta^2}}
      \,,
      ~~~~%\\
      \expval{a^\mydag a}_{n\sigma}^{(g=0)}
      =
      \omega n
      \,,
    \end{eqnarray}    
  \end{subequations}
  which are drawn in Fig.~\ref{fig1}-\ref{fig4} as dotted lines.

  The regime~\eqref{eq:validity_comm} fits the case of
  integer $M$, for which energy differences of unperturbed eigenstates
  are always integer multiples of $\omega$.  However, in the
  case of noninteger $M$, its fractional part $\delta
  M$ leads to smaller energy differences, with magnitude $|n+\sigma M
  -n'|\,\omega$, that appear in the denominator in the corrections for
  $E_{n\sigma}$. Minimizing with respect to $n'$ we find that we must
  replace~\eqref{eq:validity_comm} by the stronger requirement
  \begin{eqnarray}
    \Delta \ll \sqrt{\frac{1-|\xi|}{2}}\,\omega\lesssim g
    \,,~~~~\text{where}~
    \delta M =  M - \floor{M} = \frac{1+\xi}{2}
    \,.\label{eq:validity_inco}
  \end{eqnarray}
  For the noninteger case, our perturbation theory will thus work
  best if $|\xi|$ $=$ $0$, i.e., if $\epsilon$ $=$ $\frac14$,
  $\frac34$, $\frac54$, etc.. On the other hand, for $M$ near an integer, i.e., $\xi$ $=$
  $\pm(1-\rho)$ with $|\rho|$ $\ll$ $1$, only very small values of the
  hybridization, $\Delta$ $\ll$ $\sqrt{\rho/2}\,\omega$, can be
  expected to be accessible perturbatively.

  We use standard Rayleigh-Schr\"odinger perturbation
  theory. Nondegenerate perturbation theory applies for noninteger $M$
  (see Sec.~\ref{sec:inco-pt}) and degenerate perturbation theory for
  integer $M$ (see Sec.~\ref{sec:comm-pt}).

  \mynewpage
  \section{\label{sec:inco-pt}Perturbation theory for weak
    hybridization $\Delta$: the case of noninteger~$M$
    without degeneracies}
  
  \subsection{Eigenenergy corrections}

  As described above, the spectrum of $H_0$ is nondegenerate for
  noninteger $M$. In this case we expand the energy eigenvalue
  $E_{n\sigma}$ and eigenstate $\ket{n\sigma}$ of $H$ as a power
  series in $\dTILDE$ $=$ $\Delta/\omega$,%
  \begin{subequations}%
    \label{eq:nondegenerate_perturbation_series}%
    \begin{eqnarray}%
      E_{n\sigma}
      &=&
      E_{n\sigma}^{(0)}
      +
      E_{n\sigma}^{(1)}
      +
      E_{n\sigma}^{(2)}
      +
      \ORDER{\omega\dTILDE^3}
      \,,
      \\[1ex]
      \ket{n\sigma}
      &=&
      \ket{n\sigma^{(0)}}
      +
      \ket{n\sigma^{(1)}}
      +
      \ORDER{\dTILDE^2}
      \,.
    \end{eqnarray}
  \end{subequations}
  From the unperturbed spectrum we obtain
  \begin{subequations}%
    \begin{eqnarray}%
      \bra{n'\sigma'{}^{(0)}}V\ket{n\sigma^{(0)}}
      &=&
      \Delta
      \;
      {}_{\bar{\sigma}}\braket{n'}{n}_\sigma
      \;
      \delta_{\sigma'{}\bar{\sigma}}
      \,,
      \\
      E_{n\sigma}^{(0)}-E_{n'\sigma'}^{(0)}
      &=&
      \omega
      \,
      (n-n'+\sigma M)
      \;
      \delta_{\sigma'{}\bar{\sigma}}
      \,.
    \end{eqnarray}
  \end{subequations}
  For the first two eigenenergy corrections we obtain
  \begin{subequations}%
    \label{eq:nondegenerate_perturbed_energies}%
    \begin{eqnarray}%
      E_{n\sigma}^{(1)}
      &=&
      \bra{n\sigma^{(0)}}V\ket{n\sigma^{(0)}}
      =
      0
      \,,
      \label{eq:nondegenerate_energies_order1}
      \\[1ex]
      E_{n\sigma}^{(2)}
      &=&
      \sum_{n'\sigma'(\neq n\sigma)}
      \frac{
        |\bra{n'\sigma'{}^{(0)}}V\ket{n\sigma^{(0)}}|^2
      }{
        E_{n\sigma}^{(0)}-E_{n'\sigma'}^{(0)}
      }
      \label{eq:nondegenerate_energies_order2}
      %\nonumber
      \\\nonumber
      &=&
      \sum_{n'=0}^{\infty}
      \frac{
        \omega\,\dTILDE^2
        F_{n'n}(2\gTILDE)^2
      }{
        n-n'+\sigma M
      }
      =%\nonumber\\&=&%=
      -
      \omega\,\dTILDE^2
      \calF_n(2\gTILDE,-n-\sigma M)
      \,,\!\!
    \end{eqnarray}%
  \end{subequations}%
  where we introduced the function (for $z$ not a nonpositive
  integer) 
  \begin{eqnarray}
    \calF_n(x,z)
    &=&
    \sum_{m=0}^{\infty}
    \frac{F_{nm}(x)^2}{m+z}
    =
    e^{-x^2}\sum_{m=0}^{\infty}
    \frac{H_{nm}(x,x)^2}{n!m!(m+z)}
    \,,~~~~~~~
    z \neq 0, -1, -2, \ldots
    \,,\label{eq:calFdef}
  \end{eqnarray}
  for which a closed form and its asymptotics are derived below
  (see~\eqref{eq:calF_prettyREG_2}, \eqref{eq:calFasymp}).

  \subsection{Eigenstate corrections and expectation values}

  The first-order corrections to the 
  eigenstates are given by
  \begin{eqnarray}
    \ket{n\sigma^{(1)}}
    &=&
    \sum_{n'\sigma'(\neq n\sigma)}
    \ket{n'\sigma'{}^{(0)}}
    \frac{
      \bra{n'\sigma'{}^{(0)}}V\ket{n\sigma^{(0)}}
    }{
      E_{n\sigma}^{(0)}-E_{n'\sigma'}^{(0)}
    }
    \nonumber\\&=&%=
    \sum_{n'=0}^{\infty}
    \frac{
      \dTILDE\,
      F_{nn'}(2\gTILDE)
    }{
      n-n'+\sigma M
    }
    \,
    \ket{n'\bar{\sigma}^{(0)}}
    \,.
  \end{eqnarray}  
  Instead of normalizing these states it is preferable to obtain
  expectation values in the (normalized) perturbed eigenstates
  directly from~\eqref{eq:nondegenerate_perturbed_energies} by taking
  derivatives of the (perturbed) eigenenergy, which is given by
  \begin{subequations}%
    \label{eq:inco_expvals}
    \begin{eqnarray}%
      E_{n\sigma}(\omega,g,\epsilon,\Delta)
      &=&
      \omega n
      +
      \sigma\epsilon
      -
      \frac{\Delta^2}{\omega}
      \calF_n\Big(\frac{2g}{\omega},-n-\frac{2\sigma\epsilon}{\omega}\Big)
      +
      \ORDERBig{\frac{\Delta^3}{\omega^2}}
      \,.
    \end{eqnarray}%
    We obtain, using $X$ $=$ $2\gTILDE$ $=$ $2g/\omega$ and $Z$ $=$
    $-n-\sigma M$ $=$ $-n-2\sigma\epsilon/\omega$ for the repeatedly
    occurring arguments,
    \begin{eqnarray}%
      \expval{\NOTILDE{\sigma}_x}_{n\sigma}
      &=&
      \expval{\tz}_{n\sigma}
      =
      \frac{\partial E_{n\sigma}}{\partial\epsilon}
      =%\nonumber\\&=&%=
      \sigma
      \left(
        1
        +
        2\dTILDE^2
        \,
        \calF_n^{[0,1]}(X,Z)
      \right)
      +\ORDER{\dTILDE^3}
      ,
      \\
      \expval{\NOTILDE{\sigma}_z}_{n\sigma}
      &=&
      \expval{\tx}_{n\sigma}
      =
      \frac{\partial E_{n\sigma}}{\partial\Delta}
      =%\nonumber\\&=&
      -2\dTILDE
      \,
      \calF_n(X,Z)
      +\ORDER{\dTILDE^2}
      \\
      \expval{a^\mydag a}_{n\sigma}
      &=&
      \frac{\partial}{\partial\omega}
      \left(
        E_{n\sigma}-\frac{g^2}{\omega}
      \right)
      \\
      &=&
      n
      +
      \gTILDE^2
      +
      \dTILDE^2
      \Big[
      \calF_n(X,Z)
      +
      2\gTILDE\,
      \calF_n^{[1,0]}(X,Z)
      -
      \sigma M
      \calF_n^{[0,1]}(X,Z)
      \Big]\!
      +\ORDER{\dTILDE^3}
      \,,\nonumber
    \end{eqnarray}
  \end{subequations}
  where we use $f^{[m_1,\ldots,m_N]}(x_1,\ldots,x_N)$ $\equiv$
  $\frac{\partial^{m_1}}{\partial x_1^{m_1}}$ $\cdots$
  $\frac{\partial^{m_N}}{\partial x_N^{m_N}}$ $f(x_1,\ldots,x_N)$ as
  an abbreviation for partial derivatives throughout.
      
  \subsection{Evaluation and asymptotics}

  Here we discuss the function $\calF_n(x,z)$ in~\eqref{eq:calFdef}
  and its large-$x$ asymptotics, as well as the resulting asymptotics
  for observables depicted in Fig.~\ref{fig1}.  $\calF_n(x,z)$, as a
  function of complex $z$, has simple poles at the nonpositive
  integers.  We first consider positive real $z$ and obtain an
  explicit closed form for $\calF_n(x,z)$, which can then be
  analytically continued for all $z$ (except for the poles).
  
  For real $z$ $>$ $0$ we write
  \begin{eqnarray}
    \calF_n(x,z)
    &=&
    e^{-x^2}\int\limits_0^{1}s^{z-1}
    \sum_{m=0}^{\infty}
    \frac{H_{nm}(x,x)^2s^m}{n!m!}\,\text{d}{s}
    \label{eq:calF_as_integral}
  \end{eqnarray}
  and employ the generating function of the two-variable Hermite
  polynomials~\cite{fan_generating_2015}, specialized to equal arguments,
  %\begin{subequations}
    \begin{eqnarray}
      \sum_{m=0}^{\infty}
      \frac{H_{m,n}(x,y)H_{m,n}(x',y')s^m}{n!m!}
      % \nonumber\\&&~~
      =
      e^{sxx'}
      s^nL_{n}((\tfrac{y}{s}-x')(sx-y'))
      \,,\label{eq:H_genfunc_general}
      \\
      \sum_{m=0}^{\infty}
      \frac{H_{mn}(x,x)^2s^m}{n!m!}
      =
      e^{sx^2}s^nL_n(-x^2\tfrac{(1-s)^2}{s})
      \label{eq:H_genfunc_Laguerre}
      %\nonumber\\
      =
      e^{sx^2}
      \sum_{k=0}^{n}
      \binom{n}{k}\frac{x^{2k}}{k!}
      s^{n-k}(1-s)^{2k}
      \,,\label{eq:H_genfunc_explicit}
    \end{eqnarray}
  %\end{subequations}
  where we used the explicit form of the simple Laguerre
  polynomials $L_n(x)$ in the last line. We insert this into~\eqref{eq:calF_as_integral}
  and integrate termwise, using the integral representation of 
  the confluent hypergeometric function,
  \begin{subequations}
    \label{eq:1F1}
    \begin{eqnarray}
      \hypM(a,b,z)
      &=&
      \hypiFi(a;b;z)
      =
      \sum_{n=0}^{\infty}
      \frac{(a)_nz^n}{(b)_nn!}
      \label{eq:1F1_as_series}
      =
      e^z\hypM(b-a,b,-z)
      \\
      &=&
      \Beta(a,b-a)
      \int\limits_{0}^{1}s^{a-1}(1-s)^{b-a-1}e^{sz}\,\text{d}s
      \,,
      ~~~~\RE~b>\RE~a >0\,,
      \label{eq:_1F1_as_integral}
    \end{eqnarray}
  \end{subequations}
  where $\Beta(a,b)$ $=$ $\Gamma(a)\Gamma(b)/\Gamma(a+b)$ denotes the
  Euler Beta function.  We thus obtain a finite sum of $\hypiFi$
  functions,
  % \begin{subequations}
  %   \label{eq:calF_pretty}
    \begin{eqnarray}
      \calF_n(x,z)
      &=&
      \sum_{k=0}^{n}
      \binom{n}{k}
      \frac{x^{2k}}{k!}
      \Beta(2k+1,z+n-k)
      %\nonumber\\&&~~\times
      \hypM(2k+1,z+k+n+1,-x^2)
      %\label{eq:calF_pretty_2}
      \,,
      \label{eq:calF_pretty}
    \end{eqnarray}
  %\end{subequations}
    after using Kummer's transformation~\eqref{eq:1F1_as_series}.  The
    expressions in~\eqref{eq:calF_pretty} are also obtained if one
    instead expands the exponential in~\eqref{eq:H_genfunc_explicit}
    as a power series in $s$ and integrates termwise, using the
    integral representation of the Euler Beta function,
  \begin{eqnarray}
    \Beta(a,b)
    &=&
    =\int\limits_{0}^{1}s^{a-1}(1-s)^{b-1}\text{d}s
    \,,~~~~
    \RE~b > \RE~a > 0\,,
    \label{eq:B_as_integral}
  \end{eqnarray}
  and summing the resulting series of type~\eqref{eq:1F1_as_series}.
  Next we employ the regularized confluent hypergeometric function
  $\hypMREG(a,b,z)$, which has the advantage that it is an entire
  function of the parameters $a$ and $b$ for fixed $z$. Its definition
  and asymptotic expansion for large real argument read
  \begin{subequations}%
    \label{eq:1F1REG}
    \begin{eqnarray}%
      \hypMREG(a,b,z)
      &=&
      \frac{1}{\Gamma(b)}\hypM(a,b,z)
      =
      \sum_{n=0}^{\infty}
      \frac{(a)_nz^n}{\Gamma(b+n)\,n!}
      =
      e^z\hypMREG(b-a,b,-z)
      \label{eq:1F1REG_as_series}
      \\
      &\sim&
      e^z
      \sum_{k=0}^{\infty}
      \frac{(1-a)_k\,(b-a)_k}{\Gamma(a)\,k!\,z^{b-a+k}}
      \,,
      ~~
      z\to\infty\,~~~(a \neq 0, -1, -2, \ldots)
      \label{eq:1F1REG_asympt}
    \end{eqnarray}%
  \end{subequations}%
  The desired analytic continuation of~\eqref{eq:calF_pretty}, valid
  at all $z$ except the poles at nonpositive integers, is then given
  by
  % \begin{subequations}%
  %   \label{eq:calF_prettyREG}%
  \begin{eqnarray}
    \calF_n(x,z)
    &=&
    \sum_{k=0}^{n}
    \binom{n}{k}
    \frac{x^{2k}}{k!}
    (2k)!\,
    \Gamma(z+n-k)\,
    % \nonumber\\&&~~\times
    \hypMREG(2k+1,z+k+n+1,-x^2)
    \label{eq:calF_prettyREG_2}
    \,,
  \end{eqnarray}%
  %\end{subequations}%
  which evaluates~\eqref{eq:calFdef}. From~\eqref{eq:1F1REG_asympt} we
  find its asymptotic behavior as
  \begin{subequations}%
    \label{eq:calFasymp}%
    \begin{eqnarray}
      \calF_n(x,z)
      &\sim&
      \sum_{k=0}^{\infty}
      \;\hypiiiFii(-n,-k,k+1;1,1-z-n;1)
      %\nonumber\\&&~~~~~~~~~~\times
      \frac{(1-z-n)_k}{x^{2k+2}}
      \,,~~x^2\to\infty,
      \\[1ex]
      &=&
      \frac{1}{x^2}
      -
      \frac{(z+n-1)(z-2n)}{z\,x^4}
      +
      \ORDER{x^{-6}}
      \,.
    \end{eqnarray}%
  \end{subequations}%
  In particular, the leading order is asymptotic to $1/x^2$ and
  independent of $z$.

  Together with~\eqref{eq:inco_expvals}, this yields the following
  scenario for the large-$g$ asymptotics of $\sx$ and $\sz$: in this
  limit, due to the decay in~\eqref{eq:calFasymp} for large $x$, we
  find that $\expval{\sx}$ $\to$ $\sigma$ $=$ $\pm1$ and
  $\expval{\sz}$ $\to$ $0$ for all energy eigenstates, as in
  Fig.~\ref{fig1} (where $\epsilon$ $=$ $0.25$ is far away from 
  any half integer). As expected from the discussion in
  Sec.~\ref{sec:perturbative}, the expressions
  in~\eqref{eq:inco_expvals} describe $\expval{\sx}$ and
  $\expval{\sz}$ well only if $g$ is not too small.

  \mynewpage

  \section{\label{sec:comm-pt}Perturbation theory for weak hybridization $\Delta$: the case of integer~$M$ with degeneracies (iARM)}

  \subsection{Perturbation-diagonal eigenbasis and energy corrections}

  In the case of integer $M$ the unperturbed spectrum has
  nondegenerate and degenerate parts. For notational convenience we
  label the unperturbed energies and their eigenstates as
  $E_{n}^{(0)}$ and $\ket{n;\alpha^{(0)}}$, where $\alpha$ labels a
  possible degeneracy.

  For integer $M$ $\geq$ $1$, the lowest $M$ unperturbed energy levels
  $E_{n}^{(0)}$ and eigenstates are
  \begin{eqnarray}
    E_{n}^{(0)}
    &=&
    E_{n-}^{(0)}
    =
    \omega
    \left(n-\frac{M}{2}\right)
    \,,
    ~~
    0\leq n\leq M-1\,,
    %n=0,\ldots M-1\,,
    \\
    \ket{n;0^{(0)}}
    &=&
    \ket{n-^{(0)}}
    =
    \left(\begin{array}{cc}
        0\\[1ex]
        \ket{n}_-
      \end{array}\right)
    \,.
  \end{eqnarray}
  whose unperturbed
  eigenstates~(see~\eqref{eq:unperturbed_eigenstates}) are
  nondegenerate so that the degeneracy label $\alpha$ takes only one
  value (zero, by convention); this part of the spectrum is present
  only for nonzero $M$. The perturbation expansion for these states
  is denoted as in~\eqref{eq:nondegenerate_perturbation_series}, and
  the first-order energy correction remains zero as
  in~\eqref{eq:nondegenerate_energies_order1}.  Next follow the doubly
  degenerate unperturbed energy levels
  \begin{eqnarray}
    E_{n}^{(0)}
    &=&
    E_{n-M\hspace{0.5pt}+}^{(0)}=E_{n-}^{(0)}
    =%\nonumber\\&=&
    \omega
    \left(n-\frac{M}{2}\right)
    \,,
    ~~~~
    n=M,M+1,\ldots\,,
  \end{eqnarray}
  with unperturbed eigenstates $\ket{n-M+^{(0)}}$ and
  $\ket{n-^{(0)}}$, see~\eqref{eq:unperturbed_eigenstates}.  For the
  latter we introduce linear combinations that yield only diagonal
  matrix elements of the perturbation $V$,
  \begin{subequations}
    \begin{eqnarray}
      &\ket{n;\alpha^{(0)}}
      =
      \frac{1}{\sqrt{2}}
      \left(
        \!\!
        \begin{array}{cc}
          \ket{n-M}_+\\[1ex]
          \alpha\ket{n}_-
        \end{array}
        \!\!
      \right)\!
      ,~\alpha=\pm\,,~n\geq M,
      \\[1ex]
      &\bra{n;\alpha'^{(0)}}
      V
      \ket{n;\alpha^{(0)}}
      =\delta_{\alpha\alpha'}E_{n;\alpha}^{(1)}
      \,.
    \end{eqnarray}
  \end{subequations}
  We write the perturbative expansion for small $\Delta$ as
  \begin{subequations}
    \begin{eqnarray}
      E_{n;\alpha}
      &=&
      E_{n}^{(0)}
      +
      E_{n;\alpha}^{(1)}
      +
      E_{n;\alpha}^{(2)}
      +
      \ORDER{\omega\dTILDE^{3}}
      \,,
      \\[1ex]
      \ket{n;\alpha}
      &=&
      \ket{n;\alpha^{(0)}}
      +
      \ket{n;\alpha^{(1)}}
      +
      \ORDER{\dTILDE^{2}}
      \,.
    \end{eqnarray}
  \end{subequations}
  The degeneracies are lifted completely in first order, with
  \begin{eqnarray}%
    \label{eq:degenerate_case_energies_order1}
    E_{n;\alpha}^{(1)}
    &=&
    \bra{n;\alpha^{(0)}}V\ket{n;\alpha^{(0)}}
    \nonumber\\[1ex]
    &=&
    \begin{mycases}
      0
      &
      \text{if }n < M,~\alpha=0
      \,,
      \\[1ex]
      \alpha\omega\dTILDE F_{n,n-M}(2\gTILDE)
      &
      \text{if }n \geq M,~\alpha=\pm
      \,.
    \end{mycases}
  \end{eqnarray}
  The second-order correction reads
  \begin{eqnarray}%
    \label{eq:degenerate_case_energies_order2}%
    E_{n;\alpha}^{(2)}
    &=&
    \sum_{n'(\neq n),\alpha'}
    \frac{
      |\bra{n';\alpha'{}^{(0)}}V\ket{n;\alpha^{(0)}}|^2
    }{
      E_{n}^{(0)}-E_{n'}^{(0)}
    }
    \nonumber\\
    &=&
    -\omega\dTILDE^2
    \times
    \begin{mycases}
      \calF_n(2\gTILDE,M-n)
      &
      \text{if }n < M,~\alpha=0
      \,,
      \\[1ex]
      \tfrac12\big[
      \calG_{n}^{(n-M)}(2\gTILDE)
      +
      \calG_{n-M}^{(n)}(2\gTILDE)
      \big]
      &
      \text{if }n \geq M,~\alpha=\pm
      \,,
    \end{mycases}
  \end{eqnarray}
  where we defined $\calG_{p}^{(q)}(x)$ (for integer $p,q$ $\geq$ $0$) as
  \begin{eqnarray}
    \calG_p^{(q)}(x)
    &=&
    \sum_{m\geq0}^{(m\neq q)}
    \frac{F_{pm}(x)^2}{m-q}
    =
    e^{-x^2}
    \sum_{m\geq0}^{(m\neq q)}
    \frac{H_{pm}(x,x)^2}{p!m!(m-q)}
    \,,\label{eq:calGdef}
  \end{eqnarray}
  for which a closed form and its large-$x$ asymptotics are derived
  below (see~\eqref{eq:calGresult}, \eqref{eq:calGasymp}).
  
  \subsection{Eigenstate corrections and expectation values}

  Next we determine the first-order corrections to the
  eigenstates. For the initially nondegenerate states ($0$ $\leq$ $n$
  $<$ $M$, $\alpha$ $=$ $0$) these are
  \begin{eqnarray}%
    \ket{n;0^{(1)}}
    &=&
    \sum_{n',\alpha'}^{(n',\alpha')\neq(n,0)}
    \ket{n';\alpha'{}^{(0)}}
    \frac{
      \bra{n';\alpha'{}^{(0)}}V\ket{n;\alpha^{(0)}}
    }{
      E_{n}^{(0)}-E_{n'}^{(0)}
    }
    \nonumber\\
    &=&
    \sqrt{2}\,\dTILDE
    \sum_{n'\geq M}
    \frac{
      F_{n,n'-M}(2\gTILDE)
    }{
      n'-n
    }
    \left(\begin{array}{cc}
        \ket{n'-M}_+
        \\[1ex]
        0
      \end{array}\right)
    \,,~~~~0\leq n <M\,. 
  \end{eqnarray}%
  For the initially degenerate states ($n$ $\geq$ $M$, $\alpha$ $=$ $\pm$) we have 
  \begin{eqnarray}%
    \ket{n;\alpha^{(1)}}
    &=&
    \sum_{n',\alpha'}^{(n',\alpha')\neq(n,\alpha)}
    \ket{n';\alpha'{}^{(0)}}
    \,
    c_{n'\alpha',n\alpha}^{(1)}
    \,,
  \end{eqnarray}%
  where for different energies ($n'$ $\neq$ $n$, $\alpha'$ $=$ $\pm$)
  \begin{eqnarray}
    c_{n'\alpha',n\alpha}^{(1)}
    &=&
    \frac{
      \bra{n';\alpha'{}^{(0)}}V\ket{n;\alpha^{(0)}}
    }{
      E_{n}^{(0)}-E_{n'}^{(0)}
    }
    \nonumber\\
    &=&
    \frac{\dTILDE}{2(n-n')}\times
    \begin{mycases}
      \sqrt{2}\,F_{n',n-M}
      &
      \text{if~}0\leq n <M,~\alpha'=0
      \,,
      \\[1ex]
      \alpha'F_{n',n-M}+\alpha F_{n,n'-M}
      &
      \text{if~}n'\geq M,~n'\neq n,~\alpha'=\pm
      \,.
    \end{mycases}
  \end{eqnarray}
  while for equal energies ($\alpha'$ $\neq$ $\alpha$, hence only $\alpha'$
  $=$ $-\alpha$ $\equiv$ $\bar{\alpha}$ occurs)
  \begin{subequations}%
    \begin{eqnarray}%
      c_{n\alpha',n\alpha}^{(1)}
      &=&
      \frac{
        1
      }{
        E_{n;\alpha}^{(1)}-E_{n;\alpha'}^{(1)}
      }
      \sum_{m,\beta}^{(m\neq n)}
      \frac{
        \bra{n;\alpha'{}^{(0)}}V\ket{m;\beta^{(0)}}\,
        \bra{m;\beta{}^{(0)}}V\ket{n;\alpha^{(0)}}
      }{
        E_{n}^{(0)}-E_{m}^{(0)}
      }
      \nonumber\\
      &\equiv&
      \frac{\alpha\,\dTILDE}{2}\,
      \calC_{n,M}(2\gTILDE)\,
      \delta_{\alpha',\bar{\alpha}}
      \,,
      \\
      \calC_{n,M}(x)
      &=&
      \dfrac{
        \calG_{n}^{(n-M)}(x)
        -
        \calG_{n-M}^{(n)}(x)
      }{
        2F_{n,n-M}(x)
      }
      \,.
    \end{eqnarray}%
  \end{subequations}%
  We use these eigenstate corrections to calculate the expectation
  value of $\NOTILDE{\sigma}_x$ and  $a^\mydag a$, whereas a derivative suffices for that
  of $\NOTILDE{\sigma}_z$. We obtain
  \begin{subequations}
    \label{eq:comm_expvals}
    \begin{eqnarray}
      \expval{\NOTILDE{\sigma}_x}_{n\alpha}
      &=&
      \expval{\tz}_{n\alpha}
      =
      \begin{mycases}
        -1
        +
        \ORDER{\dTILDE^2}
        &
        \text{if }n < M,~\alpha=0
        \,,
        \\[1.5ex]
        \alpha\,\dTILDE\,
        \calC_{n,M}(2\gTILDE)\,
        +
        \ORDER{\dTILDE^2}
        &
        \text{if }n \geq M,~\alpha=\pm
        \,,
      \end{mycases}
      \label{eq:comm_sx_expval}
      \\[1ex]
      \expval{\NOTILDE{\sigma}_z}_{n\alpha}
      &=&
      \expval{\tx}_{n\alpha}
      =
      \frac{\partial E_{n;\alpha}}{\partial\Delta}
      =
      \frac{E_{n;\alpha}^{(1)}+2E_{n;\alpha}^{(2)}}{\omega\dTILDE}
      +
      \ORDER{\dTILDE^2}
      \,,
      \nonumber\\
      &=&
      \begin{mycases}
        \alpha F_{n,n-M}(2\gTILDE)
        -2\dTILDE \calF_n(2\gTILDE,M-n)
        +
        \ORDER{\dTILDE^2}
        &
        \text{if }n < M,~\alpha=0
        \,,
        \\[1ex]
        - \dTILDE \big[ \calG_{n}^{(n-M)}(2\gTILDE) + \calG_{n-M}^{(n)}(2\gTILDE) \big]
        +
        \ORDER{\dTILDE^2}
        &
        \text{if }n \geq M,~\alpha=\pm
        \,.
      \end{mycases}
      \label{eq:comm_sz_expval}
      \\[1ex]
      \expval{a^\mydag a}_{n\alpha}
      &=&
      \begin{mycases}
        n+\gTILDE^2        +
        \ORDER{\dTILDE^2}
        &
        \text{if }n < M,~\alpha=0
        \,,
        \\[1.5ex]
        n+\gTILDE^2
        -
        \dfrac{M}{2}
        \left(
          1+
          \alpha\,\dTILDE\,
          \calC_{n,M}(2\gTILDE)\,
        \right)
        +
        \ORDER{\dTILDE}
        &
        \text{if }n \geq M,~\alpha=\pm
        \,,
      \end{mycases}
      \label{eq:comm_nop_expval}
    \end{eqnarray}
    where we have calculated only part of the linear order in
    $\dTILDE$ in~\eqref{eq:comm_nop_expval}, namely that which comes
    from $c_{n\bar{\alpha},n\alpha}^{(1)}$.
    
    Note that in the integer case, i.e., for a fixed integer value
    of $2\epsilon/\omega$ $=$ $M$, the parameter $\omega$ now couples
    to the combined term $a^\mydag a$ $+$ $\frac12M\sx$ in the
    Hamiltonian, in contrast to the noninteger case of the
    previous section. From the derivative with respect to $\omega$ we
    thus obtain
    \begin{eqnarray}
      \expval{a^\mydag a}_{n\alpha}
      &+&\frac{M}{2}
      \expval{\NOTILDE{\sigma}_x}_{n\alpha}
      =
      \frac{\partial}{\partial\omega}
      \left(
        E_{n\sigma}-\frac{g^2}{\omega}
      \right)
      \nonumber\\
      &=&
      \begin{mycases}
        n+\gTILDE^2
        +
        \ORDER{\dTILDE^2}
        &
        \text{if }n < M,~\alpha=0
        \,,
        \\[1.5ex]
        n+\gTILDE^2
        -
        \dfrac{M}{2}
        +
        2\,\alpha\,\gTILDE\,\dTILDE\,
        F_{n,n-M}^{[1]}(2\gTILDE)
        +
        \ORDER{\dTILDE^2}
        &
        \text{if }n \geq M,~\alpha=\pm
        \,.
      \end{mycases}
      \label{eq:comm_nop+sx_expval}
    \end{eqnarray}
  \end{subequations}
  Note that the parts linear $\dTILDE$ listed
  in~\eqref{eq:comm_sx_expval} and~\eqref{eq:comm_nop_expval} cancel
  in~\eqref{eq:comm_nop+sx_expval}, although another linear term (from
  $\expval{a^\mydag a}$) remains in the latter.
  
  We can now understand the qualitative behavior
  of~$\expval{\sigma_x}$ as follows. Consider for now only the zeroth
  order in $\dTILDE$ in~\eqref{eq:comm_sx_expval}. In this order the
  lowest $M$ eigenstates have $\expval{\sigma_x}$ $=$ $-1$ and
  $\expval{a^\mydag a}$ $=$ $n+\gTILDE^2$, but all higher eigenstates
  have $\expval{\sigma_x}$ $=$ $0$ and $\expval{a^\mydag a}$ $=$
  $n+\gTILDE^2-M/2$. In view of~\eqref{eq:comm_nop+sx_expval} we can
  interpret this as due to the coupling of $\omega$ to both these
  expectation values for integer~$M$, and~$\expval{a^\mydag a}$
  (rather than~$\frac12M\expval{\sigma_x}$) absorbing the
  contribution~$-M/2$ for the states with $n$ $\geq$ $M$ which are
  indirectly coupled by the hybridization~$\Delta$. This provides a
  qualitative reason why, in contrast to the noninteger case,
  $|\expval{\sigma_x}|$ $\neq$ $1$ should be expected in the
  integer case. At the end of the next subsection we will discuss
  the effect of higher orders in~$\dTILDE$.

  \subsection{Evaluation and asymptotics}

  We now discuss the function $\calG_p^{(q)}(x)$
  in~\eqref{eq:calGdef}, for nonnegative integers $p$ and $q$.  First
  we express it in terms of the known function $\calF_p(x,z)$ without
  its pole $R_{pq}(x)/(z+q)$ at $z$ $=$ $-q$,
  \begin{subequations}%
    \begin{eqnarray}%
      \calG_p^{(q)}(x)
      &=&
      \lim_{z\to-q}
      \left[
        \calF_p(x,z)-\frac{R_{pq}(x)}{z+q}
      \right],
      \\
      R_{pq}(x)
      &=&
      \lim_{z\to-q}
      \left[
        (z+q)\,\calF_p(x,z)      \vphantom{\frac{R_{pq}(x)}{z+q}}
      \right]
      =
      e^{-x^2}
      \frac{H_{pq}(x,x)^2}{p!q!}
      \,.
    \end{eqnarray}%
  \end{subequations}%
  For $z$ near the pole at $-q$ we see from~\eqref{eq:calF_prettyREG_2}
  that the Gamma function yields divergent as well as regular
  contributions to $\calF_p(x,z)$. Specifically, for integer $m$ $\geq$
  $0$ and $|\delta|$ $\ll$ 1,
  \begin{eqnarray}
    \Gamma(m-q+\delta)
    =
    \begin{mycases}
      \dfrac{1}{\delta}
      \dfrac{(-1)^{q+m}}{(q-m)!}
      \Big[1+{}
      \psi(1+q-m)\,\delta+\ORDER{\delta^2}\Big]
      &\!\!
      \text{if $m$ $\leq$ $q$,}
      \\[2ex]
      (m-q)!+\ORDER{\delta}
      &\!\!
      \text{if $m$ $>$ $q$,}
    \end{mycases}
  \end{eqnarray}
  where $m$ corresponds to $n-k$ in~\eqref{eq:calF_prettyREG_2}.  In the
  second case we may set $\delta$ to zero to obtain a constant
  contribution (provided $q$ $<$ $p$), while in the first case we
  first subtract the pole $R_{pq}(x)/\delta$ and then extract the
  constant term from the next order in $\delta$ which involves
  $\hypMREG(a,b,x)$ and the Euler Digamma function $\psi(z)$ $=$
  $\Gamma'(z)/\Gamma(z)$. We thus obtain, after some rearrangement,
  \begin{eqnarray}
    \calG_p^{(q)}(x)
    &=&
    \Theta(p\!>\!q)
    \sum_{k=0}^{p-q-1}
    \binom{p}{k}
    \frac{x^{2k}}{k!}
    \,
    \Beta(2k+1,p-q-k)
    \,
    \hypM(2k+1,p-q+1+k,-x^2)
    \nonumber\\&&
    +\;
    e^{-x^2}
    \sum_{k\geq\text{max}(0,p-q)}^{p}
    \binom{p}{k}
    \frac{x^{2k}}{k!}
    \frac{(-1)^{k+p+q}(2k)!}{(k-p+q)!}
    \Big[
      \psi(k-p+q+1)\times
      \nonumber\\&&~~~~~~~~~~~~~~~~~~~~~~~~~~~~
      \hypMREG(p-q-k,p-q+1+k,x^2)
      \nonumber\\&&~~~~~~~~~~~~~~~~~~~~~~~~~~~~
      +
      \hypMREG^{[1,0,0]}(p-q-k,p-q+1+k,x^2)
      \nonumber\\&&~~~~~~~~~~~~~~~~~~~~~~~~~~~~
      +
      \hypMREG^{[0,1,0]}(p-q-k,p-q+1+k,x^2)
    \Big],
    \label{eq:calGresult}
  \end{eqnarray}%
  where the step function $\Theta(A)$ is one or zero according to whether $A$ is true
  or false, respectively; the first sum is thus absent if $q$ $\geq$
  $p$. In the second sum the hypergeometric functions terminate
  because $k$ $\geq$ $p-q$.  For large real $x$ the second sum is
  therefore exponentially small, while the asymptotic expansion of the
  first sum follows again from~\eqref{eq:1F1REG}. We obtain,
  \begin{eqnarray}%
    \label{eq:calGresult+asymp}%
    \calG_p^{(q)}(x)
    &=&
    \Theta(p\!>\!q)
    \sum_{k=0}^{\infty}
    \;\hypiiiFii(-p,-k,k+1;1,q+1-p;1)
    % \nonumber\\&&~~~~~~~~~~\times
    \frac{(q+1-p)_k}{x^{2k+2}}
    +
    \ORDER{x^{2q}e^{-x^2}}
    \nonumber\\
    &=&
    \frac{\Theta(p\!>\!q)}{x^2}
    +
    (p+q+1)
    \frac{\Theta(p\!>\!q\!+\!1)}{x^4}
    +
    \ORDER{x^{-6}}
    +
    \ORDER{x^{2q}e^{-x^2}}
    \,\label{eq:calGasymp}
  \end{eqnarray}%
  for $x^2$ $\to$ $\infty$.

  From~\eqref{eq:comm_expvals} we thus obtain the following scenario
  for the spin expectation values in Figs.~\ref{fig2}-\ref{fig4} for
  large $g$. The perturbative result for $\expval{\sz}$ tends to zero
  this limit for all energy eigenstates in agreement with the
  numerical result, as in the noninteger case. Furthermore,
  numerically we observe $\expval{\sx}$ $\to$ $-1$ for the lowest $M$
  energy eigenstates which corresponds to the (constant) perturbative
  result in linear order in $\Delta$.  In Figs.~\ref{fig2}-\ref{fig4},
  the perturbative result~\eqref{eq:comm_sx_expval} captures the
  qualitative behavior of $\expval{\sx}$ for the eigenstates with $n$
  $\geq$ $M$ in first order in $\dTILDE$ for not too small
  $g$. However, it cannot be used to obtain the large-$g$ asymptotics,
  because it contains $F_{n,n-M}(x)$ in the denominator which is
  asymptotic to $x^Me^{-x^2/2}$ and will thus eventually invalidate
  the perturbative result for large $g$. The large-$g$ asymptotics can
  thus be only partially be understood from the small-$\Delta$
  behavior in the integer case, due to the more complicated
  interplay of the expansion parameters $\gTILDE$ and $\dTILDE$.

  \mynewpage

  \section{\label{sec:parenthamiltonian}A model with number-nonconserving hybridization related to the iARM}

  \subsection{Construction in terms of exact eigenstates}

  We now construct a solvable model $H'$ which is related to the
  integer case of the asymmetric Rabi model $H$ as follows. We demand that
  it has the states $\ket{n;\alpha^{(0)}}$ (for which $V$ has only
  diagonal expectation values) as exact eigenstates but nevertheless
  contains a hybridization term $V'$, i.e.,
  \begin{subequations}%
    \label{eq:Hprime}%
    \begin{eqnarray}%
      H'
      &=&
      H_0+V'
      \,,
      \\
      H'
      \ket{n;\alpha^{(0)}}
      &=&
      E_{n\alpha}'
      \ket{n;\alpha^{(0)}}
      \,,
    \end{eqnarray}%
  \end{subequations}%
  where $H_0$ is given in~\eqref{eq:H} and $M$ is a positive integer.

  First we obtain possible forms of the operator $V'$ that are
  off-diagonal like $V$, i.e., non-commuting with $\tz$.  The
  following harmonic oscillator identities are straightforward to
  obtain from the Baker-Campbell-Hausdorff formulas,
  \begin{subequations}%
    \begin{eqnarray}%
      e^{z(a-a^\mydag)}
      \,
      (a^\mydag)^p
      &=&
      (a^\mydag+z)^p
      \,
      e^{z(a-a^\mydag)}
      \,,
      \\
      e^{2z(a-a^\mydag)}
      \ket{z}_\coh      
      &=&
      \ket{-z}_\coh
      \,,
    \end{eqnarray}%
  \end{subequations}%
  from which we obtain the operators $R_\sigma$, $\sigma$ $=$ $\pm$,
  which transform between the shifted harmonic oscillators
  of~\eqref{eq:H},
  \begin{eqnarray}
    R_{\sigma}
    \ket{n}_{\sigma}
    =
    \ket{n}_{\bar{\sigma}}
    \,,~~
    R_{\sigma}
    =
    e^{-2\sigma\gTILDE(a-a^\mydag)}
    =
    R_{\bar{\sigma}}^\mydag
    =
    R_{\bar{\sigma}}^{-1}
    \,.
  \end{eqnarray}
  We may view $R_\sigma$ as performing two successive reverse shifts $U^{-\sigma}$
  of a shifted oscillator state, one back to the original oscillator
  and one further shift into the other shifted oscillator, i.e.,
  \begin{eqnarray}
    R_{\sigma}
    &=&
    U^{-2\sigma}
    \,,~~~~
    U
    =
    e^{\gTILDE(a-a^\mydag)}
    =(U^\mydag)^{-1}
    \,.
  \end{eqnarray}
  With these operators we can transform either component of
  $\ket{n;\alpha^{(0)}}$ into the other. Namely, for $n\geq$ $M$,
  \begin{eqnarray}
    U a^M U
    \ket{n}_-
    &=&
    a_+^{M\phdag\!\!}R_-
    \ket{n}_-
    =
    w_n
    \,
    \ket{n-M}_+
    \,,
    \nonumber\\
    U^\mydag (a^\mydag)^M U^\mydag
    \ket{n-M}_+
    &=&
    R_+(a_+^\mydag)^M
    \ket{n-M}_+
    =
    w_n
    \,
    \ket{n}_-
    \,,
    \nonumber\\
    w_n
    =
    \sqrt{\frac{n!}{(n-M)!}}
    &=&
    \sqrt{n(n-1)\cdots(n-M+1)}
    \,.
  \end{eqnarray}
  so that we have ($n$ $\geq$ $M$, $\alpha$ $=$ $\pm$ 1)
  \begin{eqnarray}
    \left(\begin{array}{cc}
        0&U a^M U\\[1ex]
        U^\mydag (a^\mydag)^M U^\mydag&0
      \end{array}\right)
    \ket{n;\alpha^{(0)}}
    &=&
    \alpha
    \,
    w_n
    \,
    \ket{n;\alpha^{(0)}}
    \,.
  \end{eqnarray}
  Hence a rather general operator $V'$ with the property
  $V'\ket{n;\alpha^{(0)}}$ $=$ $v_{n\alpha}'\ket{n;\alpha^{(0)}}$ can
  be written as
  \begin{subequations}
    \label{eq:Vprime_constructed}
    \begin{eqnarray}%
      V'
      &=&
      \left(\begin{array}{cc}
          0&U\,f(a^\mydag a)\,a^M\,U\\[1ex]
          U^\mydag\,(a^\mydag)^M\,f(a^\mydag a)^*\,U^\mydag&0
        \end{array}\right),
      \\
      v_{n\alpha}'
      &=&
      \begin{mycases}
        \alpha\,w_n\,f(n-M)&n\geq M,~\alpha=\pm1
        \\
        0&0\leq n\leq M-1,~\alpha=0
      \end{mycases}
      ,\!\!\!
      \\    
      E_{n\alpha}'
      &=&
      E_{n}^{(0)}
      +
      v_{n\alpha}'
      \,,
    \end{eqnarray}%
  \end{subequations}
  where $f(n)$ is an arbitrary complex function of nonnegative integer
  $n$. For example, we might choose
  \begin{eqnarray}
    f(n)
    &=&
    \frac{\Delta_{n+M}}{w_{n+M}}
    =
    \frac{\Delta_{n+M}}{\sqrt{(n+1)\cdots(n+M)}}
    \,,
    \\
    \Rightarrow~~
    v_{n\alpha}'
    &=&
    \alpha\Delta_n
    \,.
  \end{eqnarray}
  We note a slight resemblance of $V'$ to the rotating-wave term in
  the Jaynes-Cummings Hamiltonian which also contains hermitian
  conjugate oscillator operators in the upper and lower off-diagonal.
  
  \subsection{Making the number-conserving part number-independent}

  Interestingly, $V'$ contains a number-conserving part $\bar{V}'$,
  similar to $V$ $=$ $\Delta\tx$, because $U$ is a linear combination
  of arbitrary powers of $a$ and $a^\mydag$, some of which compensate
  the factor $(a^\mydag)^M$ in $V'$. For real $f(n)$ we obtain
  \begin{subequations}
    \begin{eqnarray}
      V'
      &=&
      \widetilde{V}'
      +\text{($a^\mydag a$-nonconserving terms)}
      \\
      \widetilde{V}'
      &=&
      \tx
      \sum_{n=0}^{\infty}
      \ket{n}
      \widetilde{f}(n)
      \bra{n}
      \,,
      \\
      \widetilde{f}(n)
      &=&
      \bra{n}
      U f(a^\mydag a) a^M U
      \ket{n}
      =
      \frac{(-1)^Me^{-\gTILDE^2}}{n!}
      \sum_{m=0}^{\infty}
      \frac{f(m)}{m!}H_{n,m}(\gTILDE,\gTILDE)H_{n,m+M}(\gTILDE,\gTILDE)
      \,.\label{eq:Vbarseries}
    \end{eqnarray}
  \end{subequations}

  It is in fact possible to make these diagonal matrix elements of
  $V'$ independent of $n$, so that $\widetilde{f}(n)$ $=$ const for
  all $n$, as we now discuss. In fact, we can evaluate the series
  in~\eqref{eq:Vbarseries} for a function $f(n)$ $\propto$ $s^n$, as
  follows. 
  Since a product of two Hermite polynomials appears
  in~\eqref{eq:Vbarseries}, we first derive its generating function.
  Using the operator Hermite polynomial calculus
  of~\cite{fan_generating_2015}, we begin with
  \begin{eqnarray}
    \sum_{m=0}^{\infty}
    \frac{s^m}{m!}
    H_{m+M,n}(x,y)
    \,
    \vdots H_{m,n}(a^\mydag,a)\vdots
    &~~\stackrel{\text{\makebox[0pt][c]{[(A2)]}}}{=}~~&
    \sum_{m=0}^{\infty}
    \frac{s^m}{m!}(a^\mydag)^ma^nH_{m+M,n}(x,y)
    \nonumber\\
    &~~\stackrel{\text{\makebox[0pt][c]{[(42)]}}}{=}~~&
    e^{sa^\mydag x}H_{M,n}(x,y-sa^\mydag)a^n
    \nonumber\\[1ex]
    &~~=~~&
    H_{M,n}(x,y-sa^\mydag)(a-sx)^ne^{sa^\mydag x}
    \,.
    \label{eq:H_genfunc_shifted1}
  \end{eqnarray}
  where $\vdots\cdots\vdots$ denotes antinormal ordering of bosonic
  operators and equation numbers in square brackets refer
  to~\cite{fan_generating_2015}.  Summing over $M$ yields
  \begin{eqnarray}
    \sum_{M=0}^{\infty}
    \frac{t^M}{M!}
    H_{M,n}(x,y-sa^\mydag)(a-sx)^n
    &~~\stackrel{\text{\makebox[0pt][c]{[(21)]}}}{=}~~&
    e^{tx}(y-t-sa^\mydag)^n(a-sx)^n
    \nonumber\\[-1ex]
    &~~\stackrel{\text{\makebox[0pt][c]{[(25)]}}}{=}~~&
    e^{tx}(-s^n)H_{nn}(\I(\tfrac{y-t}{s}-a^\mydag),\I(a-sx))
    \nonumber\\[1ex]
    &~~=~~&
    e^{tx}s^nn!L_n(-(\tfrac{y-t}{s}-a^\mydag)(a-sx))
    \,.
    \label{eq:H_genfunc_shifted2}
  \end{eqnarray}
  Inside the antinormal ordering it is permissible to replace
  $a^\mydag$ and $a$ by scalars $x'$ and
  $y'$, respectively. Putting~\eqref{eq:H_genfunc_shifted1}
  and~\eqref{eq:H_genfunc_shifted2} together then
  yields
  \begin{eqnarray}
    \sum_{m=0}^{\infty}
    \sum_{M=0}^{\infty}
    \frac{H_{m+M,n}(x,y)H_{m,n}(x',y')t^Ms^m}{n!m!M!}
    % \nonumber\\&&~~
    =
    e^{sxx'+tx}
    s^nL_n((\tfrac{y-t}{s}-x')(sx-y'))
    \,.\label{eq:H_genfunc_shifted}
  \end{eqnarray}
  which is a generalization of~\eqref{eq:H_genfunc_general} that
  includes the shift $M$ in one of the indices. Next we take
  coefficients of $t^M$ in~\eqref{eq:H_genfunc_shifted}, and
  specialize to equal arguments, $x$ $=$ $y$ $=$ $x'$ $=$ $y'$,
  \begin{subequations}
    \begin{eqnarray}
      &&\sum_{m=0}^{\infty}
      \frac{H_{m+M,n}(x,y)H_{m,n}(x',y')s^m}{m!n!}
      \nonumber\\&&~~
      =
      e^{sxx'}
      \sum_{k=0}^{n}
      \binom{M}{k}
      x^{M-k}(sx-y')^ks^{n-k}L_{n-k}^{(k)}((\tfrac{y}{s}-x')(sx-y'))
      \,,
      % \end{eqnarray}
      \\
      % \begin{eqnarray}
      &&\sum_{m=0}^{\infty}
      \frac{H_{m+M,n}(x,x)H_{m,n}(x,x)s^m}{m!n!}
      \nonumber\\&&~~
      =
      e^{sx^2}x^M
      \sum_{k=0}^{n}
      \binom{M}{k}
      (s-1)^ks^{n-k}L_{n-k}^{(k)}(-x^2\tfrac{(1-s)^2}{s})
      \,,
    \end{eqnarray}
  \end{subequations}
  If we choose $f(n)$ proportional to $s^n$, with $s$ a real parameter, we thus find
  \begin{eqnarray}
    f(m)
    &=&
    \frac{s^n\Delta}{(-\tilde{g})^{M}}
    ~~\Rightarrow~~
    \widetilde{f}(m)=
    \Delta
    \sum_{k=0}^{n}
    \binom{M}{k}
    (s-1)^ks^{n-k}L_{n-k}^{(k)}(-\gTILDE^2\tfrac{(1-s)^2}{s})
    \,.
  \end{eqnarray}
  so that indeed $f(n)$ $=$ const for $s$ $=$ $1$,
  \begin{eqnarray}
    f(m)
    &=&
    \frac{\Delta}{(-\tilde{g})^{M}}
    ~~\Rightarrow~~
    \widetilde{f}(m)=
    \Delta
    ~~\Rightarrow~~
    \widetilde{V}'
    =\Delta\,\tx
    = V
    \,.\label{eq:Hprime_choice}
  \end{eqnarray}
  Remarkably, the number-conserving part $\widetilde{V}'$ in $H'$
  coincides with $\widetilde{V}$ in $H$ for the
  choice~\eqref{eq:Hprime_choice}.

  We thus arrive at the following alternative picture of the
  iARM. Namely, its Hamiltonian $H$ (with integer $M$) may be viewed
  as the number-conserving part of a parent Hamiltonian $H'$ with
  exact eigenvalues $E_{n\alpha}'$ and exact eigenstates
  $\ket{n;\alpha^{(0)}}$. The spin expectation values for $H'$ thus
  have the special $n$-dependent asymptotics described above, and
  these are unchanged in first-order perturbation theory in $(H$ $-$
  $H')$, which yields no contribution to the eigenvalues of $H$ (and by
  the derivatives in~\eqref{eq:comm_expvals} neither to the spin
  expectation values).

  Compared to the results obtained from direct perturbation theory in
  $\Delta$, this explains the $n$-dependent asymptotics of the iARM as
  due to the vicinity of the parent Hamiltonian $H'$. This picture is
  nonperturbative in $\Delta$, in the sense that the \emph{exact}
  eigenstates of $H'$ are known and agree with those of the iARM up to
  first order in $(H$ $-$ $H')$, and higher orders apparently do not
  destroy this connection.

  \mynewpage

  \section{\label{sec:conclusion}Conclusion and Outlook}

  In summary, we studied the spin expectation values in the asymmetric
  Rabi model as functions of the coupling $g$. We showed that the
  large-$g$ asymptotics can mostly be understood perturbatively for small
  hybridization $\Delta$. The spin expectation values tend to zero for
  large $g$, except in the integer case, i.e., for $\epsilon$ $=$
  $M\omega/2$ with integer $M$, for which $\expval{s_x}$ tends to $-1$
  for the $M$ lowest-lying eigenstates. As an alternative argument, we
  constructed a related Hamiltonian $H'$ with additional
  number-nonconserving terms, the exact eigenstates of which are those of
  the asymmetric Rabi model in the limit of vanishing~$\Delta$.

  As an outlook, we note that both methods hold some perspective for
  further applications.  The weak-$\Delta$ perturbation theory
  describes the crossing of energy levels on the baselines $E_n^{(0)}$
  in the integer case (left panels in
  Fig.~\ref{fig2}-\ref{fig4}), as $E_n^{(1)}$ vanishes there. This
  regime may therefore be useful to better understand the physical
  origin of these degeneracies. Similarly, the parent Hamiltonian $H'$
  captures some properties of the asymmetric Rabi model with integer 
  asymmetry parameter $M$
  and can serve as a starting point for further studies.

   \section{Acknowledgments}
   
   The authors would like to thank Daniel Braak for pointing out
   Ref.~\cite{meher_academia_2015}. Support by the Deutsche
   Forschungsgemeinschaft through TRR~80 is gratefully acknowledged.

   \bigskip\bigskip

\providecommand{\newblock}{}

\end{document}